\documentclass{techrep}
\usepackage{CFR}

\usepackage{nicefrac}
\usepackage{amsmath,amssymb}

\usepackage{fancyhdr}
\fancypagestyle{ack}{%
  \fancyhf{}

  \fancyfoot[L]{{\small This is a preprint of a chapter forthcoming in \emph{Handbook of Mixture Analysis},
    edited by Gilles Celeux, Sylvia Fr{\" u}hwirth-Schnatter, and Christian P. Robert.}}
}

\renewcommand{\thmod}{\theta}
\newcommand{\thBF}{\theta}
\renewcommand{\etav}{\eta}
\renewcommand{\Proba}[1]{\Prob ( #1)}
\newcommand{\likcom}{\llikc (\thmod, \zm;G)}
\newcommand{\ECz}{\EC (\thmod, \zm; G)}
\newcommand{\liklik}{\lik_{c} (\thmod, \zm;G)}

\newcommand{\Betadis}{\mathcal{B}e}

\newcommand{\IG}{\mathcal{IG}}

\newcommand{\Poi}{\mathcal{P}}

\DeclareMathOperator*{\argmin}{argmin}
\DeclareMathOperator*{\argmax}{argmax}

\DeclareMathOperator{\pen}{pen}

\author{Gilles Celeux\\INRIA Saclay, France \And 
  Sylvia Fr\"uhwirth-Schnatter\\Vienna University of Economics\\ and Business, Austria \And 
  Christian P. Robert\\University of Warwick,\\ UK, and Universit\'e\\ Paris-Dauphine, France}

\title{Model Selection for Mixture Models -- Perspectives and Strategies} 

\Plainauthor{G. Celeux, S. Fr\"uhwirth-Schnatter, and C.P. Robert}

\begin{document}
\maketitle
\thispagestyle{ack}

\vspace{-0.8cm}
\tableofcontents
\vspace{0.4cm}


\section{Introduction}

Determining the number $G$ of components in a finite mixture distribution  defined as\index{number of components}
\begin{align}
  y \sim  \sum \limits_{g=1}^G \eta_g f_g(y |\theta_g), \label{eq:mixOfMix1}
\end{align}
is an important and difficult issue.
This is a most important question, because statistical inference about the resulting
model is highly sensitive to the value of $G$. Selecting an erroneous value of $G$ may produce a poor density
estimate. This is also a most difficult question from a theoretical perspective
as it relates to unidentifiability issues of the mixture
model,\index{identifiability} as discussed already in Chapter~4.
This is 
a most relevant question from a practical viewpoint since the
meaning of the number of components $G$ is strongly related to the modelling
purpose of a mixture distribution.

From this perspective, the famous quote from\index{Box, George} \citet{Box},
\lq\lq All models are wrong, but some are useful\rq\rq\,  is particularly
relevant for mixture models since they may be viewed as a semi-parametric tool
when addressing the general purpose of density estimation or as a model-based
clustering tool when concerned with unsupervised classification; see also
Chapter~1. Thus, it is highly desirable and ultimately profitable
to take into account the grand modelling purpose of the statistical analysis when selecting a proper value of
$G$, and we distinguish in this chapter between selecting $G$ as a density estimation problem in\index{mixtures!as density estimates}
Section~\ref{sec_dens}  and selecting $G$ 
 in  a model-based clustering framework\index{mixtures!as clustering tool}
 in Section~\ref{secG_clust}.

Both sections will discuss frequentist as well as Bayesian approaches. At a foundational level,
the Bayesian approach is often characterized as being highly directive, once the prior distribution has been chosen
\citep[see, for example,][]{robert:2007}. While the impact of the prior on the evaluation of the number   of components in a
mixture model or of the number   of clusters in a sample from a mixture distribution cannot be denied, there exist competing ways of assessing these quantities, some borrowing from point estimation and others from hypothesis
testing or model choice, which implies that the solution produced will strongly
depend on the perspective adopted. We present here some of the Bayesian\index{Bayesian!foundations}\index{Bayesian!inference}
solutions to the different interpretations of picking the \lq\lq right\rq\rq\  number of components in a mixture,
before concluding on the ill-posed nature of the question.\index{ill-posed problem}

As already mentioned in  Chapter~1, there exists an intrinsic and foundational difference between frequentist and Bayesian inferences:
only Bayesians can truly {\em estimate} $G$, that is, treat $G$ as an additional
unknown parameter that can be estimated simultaneously with the other model\index{number of components!as extra parameter}
parameters $\thmod=(\eta_1,\ldots,\eta_G, \theta_1,\ldots,\theta_G)$ defining the mixture distribution (\ref{eq:mixOfMix1}). Nevertheless, Bayesians very often rely on model\index{model selection!Bayesian}
selection perspectives for $G$, meaning that Bayesian inference is carried out for a range of
values of $G$, from $1$, say, to a pre-specified maximum value $\Gmax$, given a sample $\ym=(y_1, \ldots,y_n)$ from (\ref{eq:mixOfMix1}). Each value
 of $G$ thus corresponds to a
potential model ${\cal M}_G$, and those models are compared via Bayesian model selection.  A typical choice for conducting this
comparison is through the values of the marginal likelihood $p(\ym| G)$,\index{marginal likelihood}
\begin{equation}\label{eq:marlik}
p(\ym | G)=\int p(\ym| \thmod, G) p(\thmod |G ) \text{d} \thmod,
\end{equation}
separately for each mixture model  ${\cal M}_G$, with $p(\thmod |G )$ being a prior distribution for all unknown parameters $\thmod$ in
a mixture model with $G$ components.

However, cross-model Bayesian inference on $G$ is far more attractive, at least
conceptually, as it relies on one-sweep algorithms, namely computational
procedures that yield estimators of $G$ jointly with the unknown model
parameters.   Section~\ref{sec_bayes} reviews such one-sweep Bayesian
methods for cross-model inference on $G$, ranging from well-known methods such
as reversible jump Markov chain Monte Carlo (MCMC) to more recent ideas involving sparse finite mixtures
relying on overfitting in combination with a prior on the
weight\index{cross-model Bayesian inference}\index{RJMCMC}\index{reversible jump MCMC}\index{MCMC!reversible jump}\index{sparsity} distribution that forces sparsity.

 \begin{figure}[t!]
\begin{center}
\scalebox{0.3}{\includegraphics{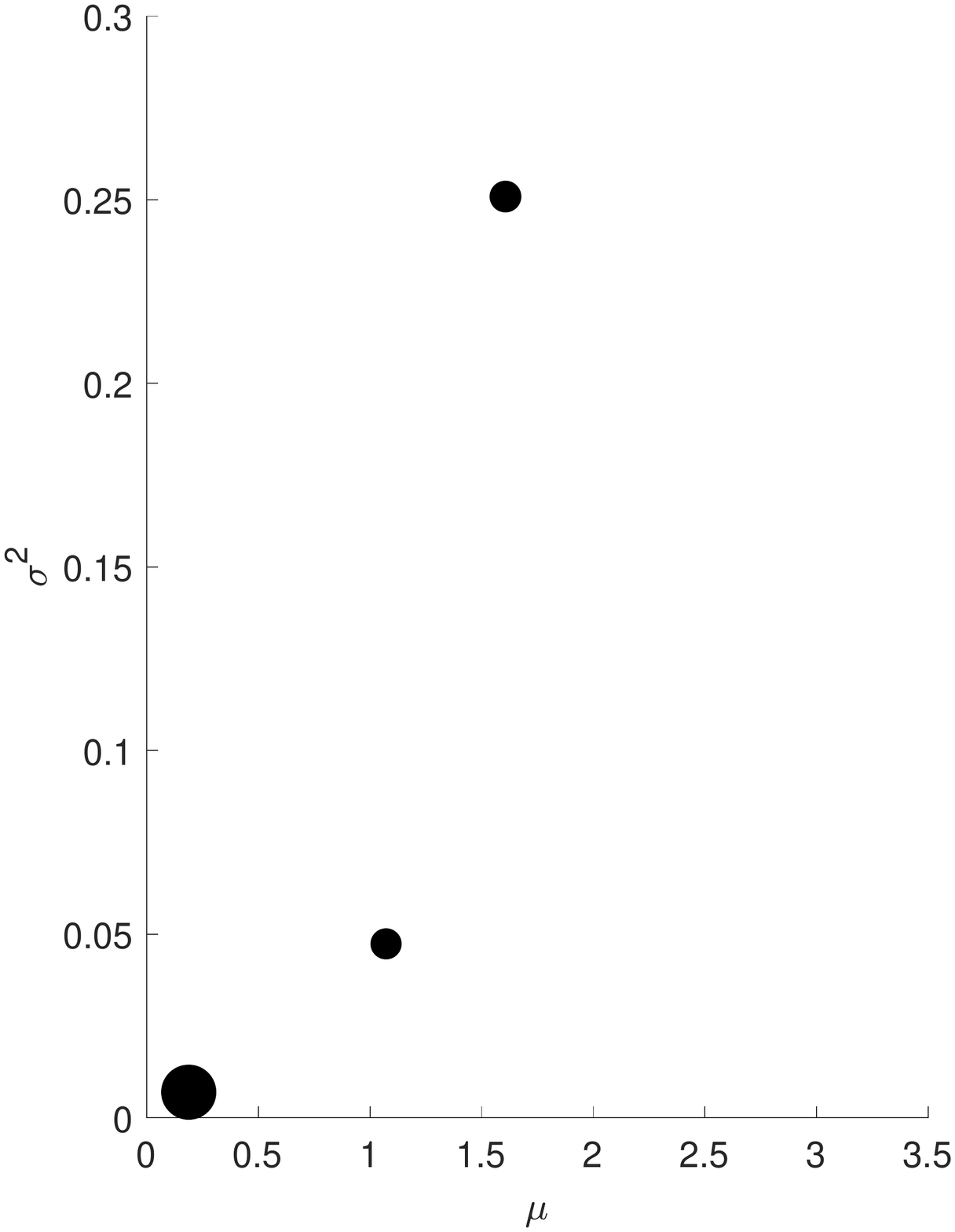}}\\[2mm]
\scalebox{0.3}{\includegraphics{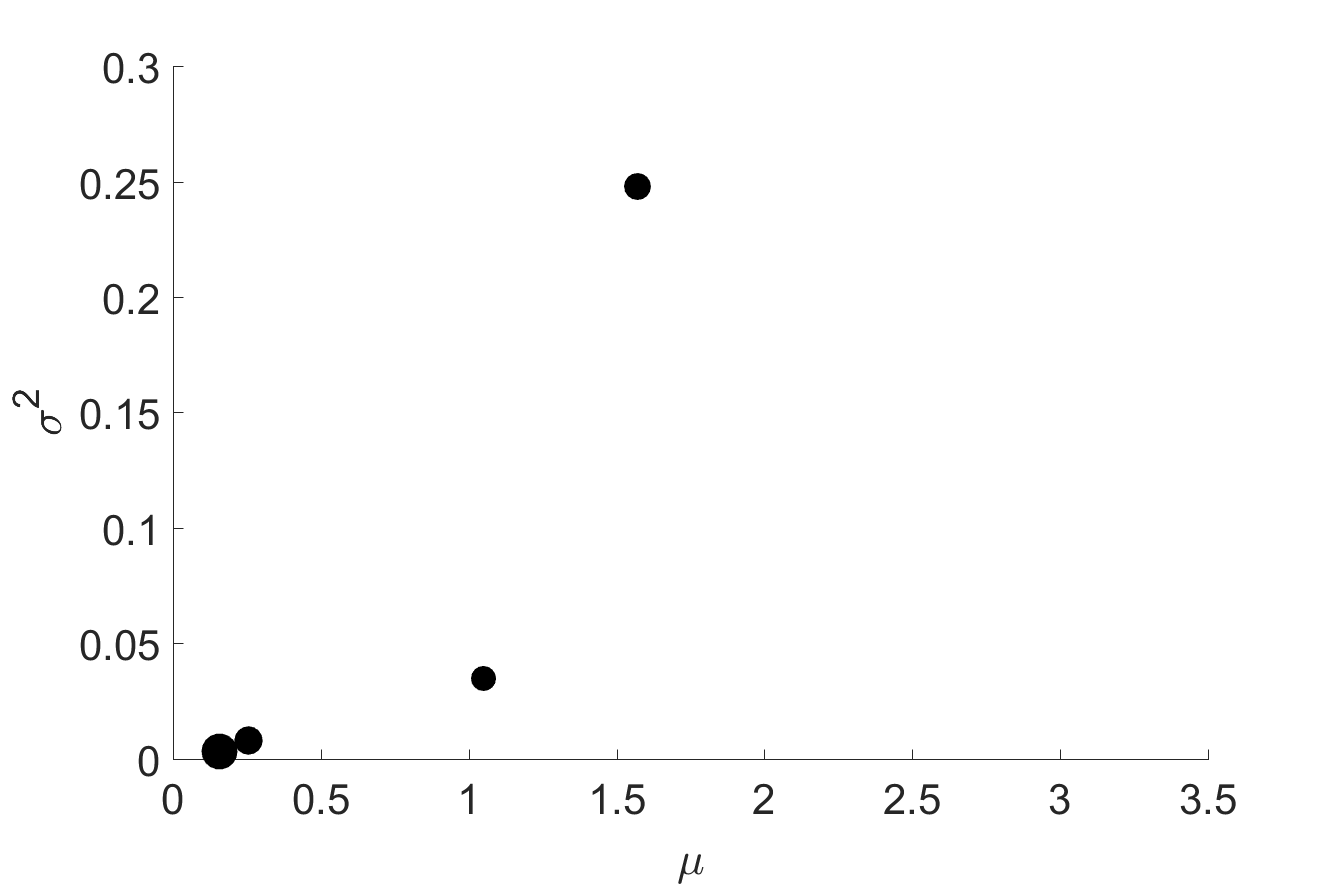}} \scalebox{0.3}{\includegraphics{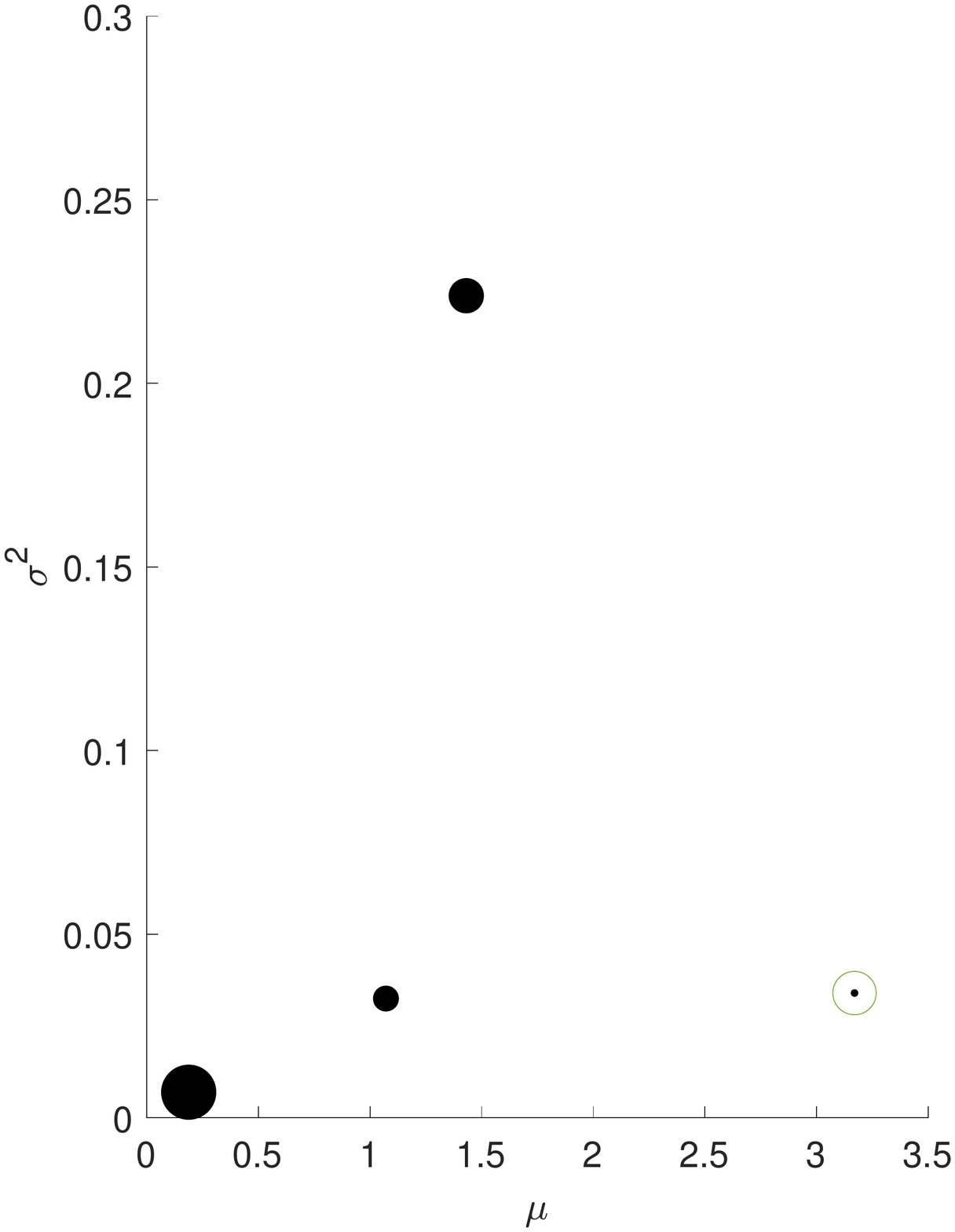}}
\end{center}
\caption{Point process representation of  the estimated mixture parameters  $(\hat{\mu}_g, \hat\sigma_g^2)$  
for three  mixture distributions  fitted to  the enzyme data
using a Bayesian framework under the prior of \citet{richardson:green:1997}. The size of each point 
corresponds to the mixture weight $\hat\eta_g$.
Top: $G=3$. Bottom:  $G=4$  with  $\eta \sim \Dirinv{4}{4}$ (left)
 and  $\eta \sim \Dirinv{4}{0.5}$ (right; the very small fourth component is marked by a circle).}\label{cfr:fig2}
\end{figure}


\section{Selecting $G$ as a Density Estimation Problem} \label{sec_dens}

When the estimation of the data distribution is the main purpose of the mixture modelling, it is generally assumed that this
distribution truly is a finite mixture distribution. One inference issue is then to find the true number of mixture components,\index{number of components!true value of}\index{order}
$G$, that is, the {\em order} of the mixture behind the observations. This assumption is supposed to produce well-grounded
tests and model selection criteria.

The {\em true order} of a finite mixture model is the smallest value of $G$ such that the components of the mixture in (\ref{eq:mixOfMix1}) are
all distinct   and the mixing proportions are all positive (that is, $\theta_g \neq  \theta_{g'}$, $g\neq  g'$ and $\eta_g>0$).  This definition
attempts to deal with the  ambiguity (or non-identifiability) due to overfitting, discussed\index{identifiability}
in  Chapter~1 and Chapter~4: 
a mixture with $G$ components can equally be defined as a (non-identifiable) mixture with $G+1$ components where the additional component
either has a mixing proportion $\eta_{G+1}$ equal to zero or the  parameter $\theta_{G+1}$ is identical to the parameter $\theta_{g}$ of some
 other component $g \in \{1, \ldots, G\}$.
 These identifiability issues impact both frequentist and Bayesian methods for selecting $G$.
 Hence, the order $G$  is a poorly
 defined quantity and  in practical mixture analysis   it is  often  difficult to decide what order $G$  describes the data best.

 \begin{figure}[t!]
\begin{center}
\scalebox{0.6}{\includegraphics{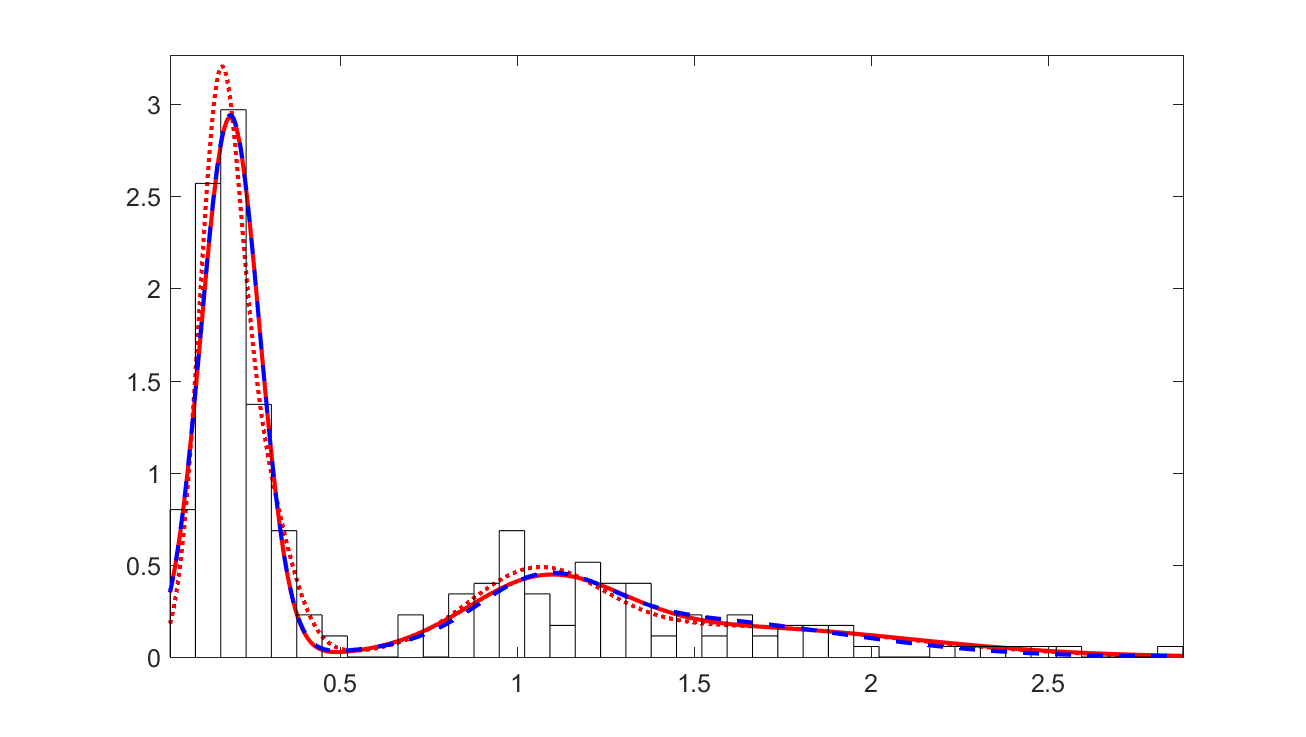}}
\end{center}
\caption{Histogram of the enzyme data together with three fitted mixture distributions:
$G=3$ (solid line);     $G=4$  and  $\eta \sim \Dirinv{4}{4}$ (dotted line);    $G=4$  and  $\eta \sim \Dirinv{4}{0.5}$ (dashed line).
The dashed and solid lines are nearly identical.}\label{cfr:fig1}
\end{figure} 

By way of illustration,  a  mixture of normal  distributions $\Normal(\mu_g, \sigma_g^2)$ with $G=3$ components is fitted within a Bayesian framework to the enzyme data studied in \citet{richardson:green:1997}, using the same prior as \citeauthor{richardson:green:1997}, in particular a uniform prior on the  weight distribution $\eta=(\eta_1, \ldots, \eta_G)$. In addition,
 mixtures with  $G=4$  components are fitted, but with
 different symmetric Dirichlet priors for $\eta$, namely $\eta \sim \Dirinv{4}{4}$  and  $\eta \sim \Dirinv{4}{0.5}$.
As discussed in Chapter~4 above,
the first prior  favours overlapping  components, whereas the second prior
favours small components, should the mixture be overfitting.

Full conditional Gibbs sampling is applied for  posterior inference.
  All three mixture models are identified  by $k$-means clustering in the point process representation of the
 posterior draws of $(\mu_g,\sigma_g)$.  The estimated  component parameters
  $(\hat{\mu}_g, \hat{\sigma}_g^2,\hat{\eta}_g)$ are visualized through a point process representation in Figure~\ref{cfr:fig2}.
  Obviously, the parameters for the four-component mixture are quite different and emerge in quite different ways than the components
  of the three-component mixture.
 The component   $(\hat{\mu}_g, \hat{\sigma}_g^2,\eta_g)=(0.19,0.007,0.61)$
  is split  into the two components  $(\hat{\mu}_g, \hat{\sigma}_g^2)=(0.16, 0.003)$ and
 $(\hat{\mu}_{g'}, \hat{\sigma}_{g'}^2)=(0.26,0.008)$
 with     weights  
 $0.38+ 0.23 =0.61$ under the  prior $\eta \sim \Dirinv{4}{4}$.
Under the prior  $\eta \sim \Dirinv{4}{0.5}$,  the variance of the two components with the larger means is  reduced
and a fourth  tiny component with  weight $0.012$ and a large
mean 
is added.


 Figure~\ref{cfr:fig1} shows the density of  these three mixture distributions together
 with a  histogram of  the data. The density of $G=4$ under the prior $\eta \sim \Dirinv{4}{0.5}$ is nearly identical to
    the density of $G=3$ with the tiny fourth component  capturing the largest observations.
The density of $G=4$ under the prior $\eta \sim \Dirinv{4}{4}$   is also very similar to
    the density of $G=3$, but  tries to capture the skewness  in the large, well-separated  cluster with the smallest observations.
 Clearly, it is not easy to decide  which of these three densities describes the data best.

\subsection{Testing the order of a finite mixture through likelihood ratio tests} \label{iden}

From a frequentist perspective, a natural approach to the determination of the
order of a mixture distribution is to rely on the likelihood ratio test\index{likelihood!ratio test}\index{test!likelihood ratio}
associated with the hypotheses of $G$ ($H_0$) versus $G+1$ ($H_A$) non-empty
components.  However, as a consequence of the above-mentioned identifiability
problem,  regularity conditions ensuring a standard asymptotic distribution for
the maximum likelihood (ML) estimates do not hold; see Chapter~4.
When one component is superfluous ($H_0$), the
parameter $\thmod_{G+1}$ under the alternative hypo\-thesis $H_A$ lies on  the
boundary of the parameter space. Moreover, the remainder term appearing within
a series expansion of the likelihood ratio test statistic is not uniformly
bounded under  $H_A$. Therefore,  its distribution remains unknown.

Many
attempts have been made to modify the likelihood ratio test in this setting; see, for example, the references in \citet{McLPeel} and \citet{fruhwirth:2006}.
Here, we wish to mention the seminal works of \citet{dacunha:gassiat:1997b,dacunha:gassiat:1999},
which make use of a locally conic parameterization to deal with non-identifiability.\index{parameterization!conic}
This research has been
updated and extended to ensure a consistent estimation of $G$ with penalized ML
when $G$ is bounded for independent and dependent finite mixtures \citep{gassiat:2002}.
Note that this boundary on $G$ has been relaxed in the paper of \citet{GassiatHandel} for a mixture of translated distributions.
Moreover, an early reference that deals explicitly with testing $G$ against $G+1$ in Markov switching models (see Chapter~11) is \citet{Hansen92}.

Adopting a different perspective, \citet{MCLachlan87} proposed using a parametric bootstrap test to select the number of\index{test!bootstrap}
components in a normal mixture. This approach can be extended without difficulty to other mixture  distributions.
To test the null hypothesis  that $G=G_0$ against the alternative that $G=G_1$   at the level $\alpha$,
\citet{MCLachlan87} suggests the following procedure: draw $B$ bootstrap
samples from a  mixture model of order $G_0$ with the parameters being equal to
the maximum likelihood estimator (MLE) $\hat \thmod_{G_0}$ and compute the log likelihood ratio
statistic (LRS) of  $G=G_0$ versus $G=G_1$ for all bootstrap samples. If the
LRS computed on the original sample is smaller than the $1-\alpha$ quantile
of the distribution of the bootstrapped LRSs,  then the hypothesis   $G=G_0$
is not rejected.  It must be pointed out that this bootstrap test is {\em
biased} since the $p$-value is computed from a bootstrap sample where the
parameter value $\thmod_{G_0}$ has been estimated from the {\em whole} observed
sample. One way to address 
this bias is to resort to {\em double\index{bootstrap!double}
bootstrapping}: first, $B$ bootstrap samples are used to compute an  estimate
$\hat \thmod_{G_0}^b$ for each bootstrap sample $b=1, \ldots, B$, while a
second bootstrap layer produces an LRS for each bootstrap sample $b$ of the
first bootstrap layer. Unfortunately, this double bootstrap procedure is
extremely computer-intensive.

As far as we know, technical difficulties aside, statistical tests are rarely used to estimate the order
of a mixture.  There are several reasons for this. First, the mixture models under comparison are not necessarily embedded.
And second, the proposed tests are numerically difficult to implement and slow. Hence, other procedures such as optimizing
penalized log likelihood or resorting to Bayesian methods are preferable.

\subsection{Information criteria for order selection}  \label{sec:infcri}

 Various  information criteria 
 for selecting the order
 of a  mixture distribution  are discussed in this section, including the Akaike (AIC) and Bayesian (BIC) information criteria
(Section~\ref{aic_bic}), the slope heuristic (Section~\ref{slope}), the deviance information criterion (DIC)
(Section~\ref{pen_log_lik}), and the minimum message length (Section~\ref{cgs:mimlen}) and we refer to the literature
for additional criteria such as the approximate weight of evidence (AWE) criterion \citep{ban-raf:mod}.\index{penalty}\index{likelihood!penalized}
 Information criteria are based on penalizing   the log  likelihood function  of a mixture model ${\cal M}_G$ with $G$ components,
$\lliko (\thmod;G) = \log \, \liko (\thmod;G)$, where
\begin{equation}  \label{deffLO}
\liko (\thmod;G) =\prod_{i=1}^{n} \left[\sum_{g=1}^{G} \eta_{g}f_g(y_{i}
\mid {\mathbf \theta}_{g})\right]
\end{equation}
is  also known as the observed-data likelihood.
The penalty  is proportional to the number of free parameters in  ${\cal M}_G$, denoted by $\upsilon_G$,
and the various criteria differ in the choice of the corresponding proportionality factor.
 The number $\upsilon_G$ increases linearly  in $G$  and  quantifies the complexity of the model. For a multivariate mixture of Gaussian distributions with unconstrained
 covariance matrices generating observations  of dimension $r$, for instance, $\upsilon_G= G(1+r + r(r+1)/2)-1$.

\subsubsection{AIC and BIC}  \label{aic_bic}

Let $\hat\thmod_G$  be the  MLE  corresponding  to the observed-data likelihood $\liko (\thmod;G)$, defined
in (\ref{deffLO}). The
AIC \citep{Akaike} and BIC \citep{Schwarz} are popular model selection criteria for solving the bias--variance\index{AIC}\index{Akaike information criterion (AIC)}\index{BIC}\index{Bayesian information criterion (BIC)}
dilemma for choosing a parsimonious model.
 AIC$(G)$ is defined as
\begin{equation}
\mbox{AIC}(G)= -2 \,  \lliko (\hat\thmod_G;G) + 2 \, \upsilon_G,
\end{equation}
whereas BIC$(G)$ is defined as
\begin{equation}
\mbox{BIC}(G)= -2 \, \lliko (\hat\thmod_G;G) + \upsilon_G \log(n).
\end{equation}
Both criteria are asymptotic criteria and assume that the sampling pdf is
within the model collection. On the one hand, the AIC aims to minimize the Kullback--Leibler divergence\index{Kullback--Leibler divergence} between model ${\cal M}_G$ and the
sampling pdf. On the other hand, the BIC approximates the marginal likelihood of model ${\cal M}_G$, defined in
(\ref{eq:marlik}), by ignoring the impact of the prior.\index{marginal likelihood}

In some settings and under proper regularity conditions, the BIC can be shown to be consistent, meaning it eventually picks
the true order of the mixture, while the AIC is expected to have a good predictive behaviour and happens to be minimax optimal, that is, to minimize
the  maximum risk among all estimators, in some regular situations \citep{Yang05}. However, in a mixture setting both
penalized log likelihood criteria face the same difficulties as the likelihood ratio test due to the identifiability problems mentioned in Section \ref{iden}.

Under proper regularity conditions, the BIC enjoys the following asymptotic properties.
\begin{itemize}
\item[(a)] The BIC is consistent:\index{BIC!consistency}\index{Bayesian information criterion (BIC)!consistency} if there exists $G^*$ such that the true distribution $p_0$ generating the data
is equal to $p(\cdot|G^*)$,
then, for $n$ large enough, BIC selects $G^*$.
\item[(b)] Even if such a $G^*$ does not exist,
 good behaviour of the BIC can be expected,  if $p_0$ is close to $ p(\cdot|G^*)$ for the value $G^*$ selected  by the BIC.
\end{itemize}
Unfortunately, the regularity conditions that validate the above Laplace approximation require the model parameters to be identifiable.
As seen above, this is not true in general for most mixture models.  However, the BIC has been shown to be consistent when
the pdfs of the mixture components are bounded \citep{Keribin}. This is, for example,  the case for a Gaussian mixture
model with equal covariance matrices. In practice, there is no reason to think that the BIC is not consistent for selecting
the number of mixture components when the mixture model is used to estimate a density \citep[see, for instance,][]{Roeder97,Fraley02}.

For singular models for which the Fisher information matrix is not everywhere invertible,
\cite{DrtonPlummer2017} proposed the so-called sBIC criterion.\index{sBIC} This criterion makes use of the \cite{Watanabe2009} marginal
 likelihood approximation of a singular model.\index{marginal likelihood}
It is the solution of a fixed point equation approximating the weighted average of the log marginal likelihoods of the models in competition.
The  sBIC criterion is proven to be consistent. It coincides with the BIC criterion  when the model is regular. But,
while the usual BIC is in fact not Bayesian, the
sBIC is connected to the large-sample behaviour of the log marginal likelihood  \citep{DrtonPlummer2017}.

However, the BIC does not lead to a prediction of the observations that is asymptotically optimal; see \citet{Yang05} and \citet{DrtonPlummer2017}
for further  discussion on the comparative  properties of the AIC and BIC.
In contrast to the BIC criterion, the AIC is known to suffer from a marked tendency to overestimate the true value of $G$
(see, for instance, \citet{Celeux96} for illustrations).  However, a modification of AIC, the so-called AIC3 criterion\index{AIC3},
proposed in \cite{Bozdogan87}, which replaces the penalty $2\upsilon_G$ with $3\upsilon_G$,  provides a good assessment of
$G$ when the latent class model is used to estimate the density of categorical data \citep{Nadif}. Nevertheless,
the theoretical reasons for this interesting behaviour of the AIC3 (in this particular context) remain for the most part mysterious.

Finally, when the BIC is used to select the number of a mixture components for real data, it has a marked tendency to
choose a large number of components or even to choose the highest proposed number of components. The reason
for this behaviour is once more related to the fact that the penalty of the BIC is independent of the data, apart from the sample size $n$. When the bias
in the mixture model does not vanish when the number of components increases, the BIC always increases by
adding new mixture components. In a model-based clustering context, this under-penalization tendency
is often counterbalanced by the entropy of the mixture, added to the BIC in the ICLbic criterion  (see Section \ref{icl}), which could lead to
a compromise between the fit of a mixture model and its ability to produce a sensible clustering of the data. But there are many
situations where the entropy of the  mixture
is not enough for counterbalancing this tendency and, moreover, the ICLbic is not really relevant when the modelling purpose
is not related to clustering.

\subsubsection{The Slope Heuristics}  \label{slope}
The so-called slope heuristics \citep{BirgeMassart01,BirgeMassart07},\index{slope heuristics}
are a data-driven method to calibrate a penalized criterion that is known up to a multiplicative constant $\kappa$.
It has been successfully applied to many situations, and particularly to mixture
models when using the observed-data  log likelihood; see \cite{BMM12}.
As shown by \cite{Baudry15}, it can be
extended without difficulty to other contrasts including the conditional\index{likelihood!penalized}
 classification  log likelihood,  which will be defined  in
Section \ref{icl}.  Roughly speaking, as with the AIC and BIC, the penalty function
$\mbox{pen}(G)$ is assumed to be proportional to
the number of free parameters $\upsilon_G$ (i.e.\ the model dimension), $\mbox{pen}(G) \propto \kappa \upsilon_G$.

The penalty is calibrated using the data-driven slope estimation (DDSE) procedure, available in the R package {\sf capushe}
\citep{BMM12}. The method assumes a linear relation between the observed-data
 log likelihood and the penalty.\index{R package!capushe@{\sf capushe}}
It is important to note that this assumption must and may easily be verified in practice via a simple plot.
Then the DDSE procedure directly estimates the slope
of the expected linear relationship  between  the contrast (here the observed-data log likelihood, but other contrasts
such  as the conditional classification likelihood are possible)  and  the model dimension  $\upsilon_G$
 which is a function of the number $G$ of components.
The estimated slope $\kappa$ defines a minimal penalty $\kappa \upsilon_G$ below  which smaller penalties give rise
to the selection of   more complex models, while higher penalties should select models with reasonable complexity.
Arguments are provided  in \citet{BirgeMassart07} and  \citet{BMM12}  that  the optimal (oracle) penalty is  approximately
 twice the minimal penalty. Thus, by setting the penalty to be $ 2 \kappa \upsilon_G$,   the slope
 heuristics criterion is defined  as\index{oracle!penalty}\index{penalty!oracle optimal}
\[
\mbox{SH}(G)= -\lliko (\hat\thmod_G;G) + 2  \kappa  \upsilon_G,
\]
when considering  mixture models in a density estimation framework.
For more details  about the rationale and the implementation of the slope heuristics, see \citet{BMM12}.

The slope heuristics method relies on the assumption that the bias of  the  fitted models
decreases as their complexity increases and becomes almost constant
for the most complex model. In the mixture model framework, this requires the family of models to be roughly
nested. More discussion, technical developments and illustrations are given in \citet{Baudry15}.

The ability of the slope heuristics method, which is not based on asymptotic arguments,
to detect the stationarity of the model family bias 
(namely the fact that the bias becomes almost constant) is of prime relevance.
It leads this criterion to propose more parsimonious models than the BIC or even the integrated complete-data likelihood criterion (to be discussed in Section \ref{icl}). Many illustrations of this practical
behaviour can be exhibited in various domains of application of  mixture models; see, for instance, a clustering use of the slope heuristics to
choose the number of components of a multivariate Poisson mixture with RNASeq transcriptome data \citep{RMMC15}
or in a model-based clustering approach for comparing  bike sharing systems  \citep{BCJ15}.

\subsubsection{DIC}  \label{pen_log_lik}


In recent years, the deviance information criterion\index{deviance information criterion (DIC)} 
introduced by \citet{spi-etal:baydic} has become a popular
criterion for Bayesian model selection because it is easily computed from posterior draws, using MCMC 
methods.
  Like other penalized log likelihood criteria, the DIC
involves a trade-off between goodness of fit and model complexity,
measured  in terms of the so-called effective number of parameters.
However, the use of the DIC to choose the order $G$  of a  mixture model  is not
without issues, as discussed by \citet{deiorio:robert:2002} and \citet{cel-etal:dev}.

To apply the DIC  in a mixture context, several decisions have  to be made.
As for any latent variable model,  a first difficulty  arises in the choice of the appropriate likelihood\index{likelihood!observed-data}\index{likelihood!conditional}\index{likelihood!complete-data}
function. Should the DIC be based on  the {\em observed-data} log  likelihood  $\log\, p(\ym|\thmod, G)$,  the {\em complete-data} log likelihood
$\log\, p(\ym, \zm|\thmod, G)$  or the {\em conditional} log  likelihood  $\log\, p(\ym| \zm,\thmod, G)$, where  $ \zm=(z_1, \ldots,z_n)$
are the latent allocations generating the data (see also Section~\ref{sec:exp_par})?
Second, the calculation of the DIC requires an estimate $\hat{\thmod}_G$ of the unknown parameter $\thmod$ which may suffer from label
switching, making the DIC (which is based on averaging over MCMC draws)  unstable.  Finally,  if the definition of the DIC  involves either the complete-data
or conditional likelihood, the difficulty that $\zm$ is unobserved  must be dealt with, either by
integrating against the posterior  $p(\zm|\ym,G)$ or by using a plug-in estimator of $\zm$ in which  case  once again the
label switching problem must be addressed to avoid instability.

In an attempt to calibrate these difficulties, \citet{cel-etal:dev} investigate in total eight different DIC criteria.
DIC$_2$, for instance, focuses on the marginal distribution of the data and considers the allocations $\zm$ as nuisance
parameters. Consequently, it is based on the observed-data likelihood:
$$
\mbox{DIC}_{2} (G)  =  -4 \Ethdis{\thmod}{\log\, p(\ym|\thmod,G)|\ym} +
  2 \log\, p(\ym|\hat{\thmod}_G, G), \label{DIC2}
$$
where  the posterior mode estimator $\hat{\thmod}_G$ (which is invariant to label switching\index{label switching}) is obtained from the
observed-data posterior $p(\thmod|\ym,G)$ and $\Ew_\thmod$ is the expectation with respect to the posterior
$p(\thmod|\ym,G)$.

Based on several simulation studies, \citet{cel-etal:dev} recommend using DIC$_4$ which is based on computing first DIC
for the complete-data likelihood function and then integrating over $\zm$ with respect to the posterior $p(\zm|\ym,G)$.
This yields
$$%
\mbox{DIC}_{4} (G)  =  -4 \Ethdis{\thmod,\zm}{\log\, p(\ym, \zm|\thmod, G)|\ym} +
  2 \Ethdis{\zm}{\log\, p(\ym, \zm|\hat{\thmod}_G(\zm))|\ym}, \label{DIC4}
$$%
where  $\hat{\thmod}_G(\zm)$ is  the complete-data posterior mode which must be
computed   for each
draw from the posterior $p(\zm|\ym,G)$.  This is straightforward if  the complete-data posterior
$p(\theta_g|\ym, \zm)$ is available in closed form.  If this is not the case,   \citet{cel-etal:dev}
instead use the posterior mode estimator $\hat{\thmod}_G$ of the observed-data posterior $p(\thmod|\ym)$.
 This  leads to  an approximation of $\mbox{DIC}_{4} (G)$,
called $ \mbox{DIC}_{4a}(G) $,  which is  shown to be a criterion that penalizes $\mbox{DIC}_{2} (G)$  by the expected entropy, defined in (\ref{ent}):
$$
\mbox{DIC}_{4a}  (G) = \mbox{DIC}_{2} (G) +  2 \Ethdis{\thmod}{\ENT ( \thmod; G) |\ym}.
$$
Both $\mbox{DIC}_{2} (G) $ and $\mbox{DIC}_{4a} (G) $ are easily estimated from  (MCMC) draws from  the posterior
$p(\thmod|\ym,G)$  by substituting all expectations $\Eth{\bullet}{\cdot|\ym}$ by an average over the corresponding draws. Note
that label switching is not a problem here, because both $\log\, p(\ym|\thmod, G)$ and $\ENT ( \thmod; G)$ are invariant
to the labelling of the groups.

However,  in practical mixture modelling, the DIC turns out to be very unstable, as shown by \citet{cel-etal:dev} for the galaxy data \citep{roe:den}.
A similar behaviour was observed by  \citet{fru-pyn:bay} 
who fitted skew-normal mixtures to Alzheimer disease data  under various prior
assumptions. While the marginal likelihood selected $G=2$ with high confidence for all priors,
 $\mbox{DIC}_{4a} (G)$ selected $G=1$,  regardless of the chosen prior, whereas the number of components selected
 by $\mbox{DIC}_{2} (G)$  ranged  from 2 to 4, depending on the  prior.

\subsubsection{The minimum message length}   \label{cgs:mimlen}

Assuming that the form of the mixture models is fixed (e.g.\ Gaussian mixture models with free covariance matrices
or Gaussian mixture models with a common covariance matrix), several authors have proposed dealing with the estimation of the mixture para\-meters and $G$\index{MML}\index{minimum message length (MML)}
in a single algorithm with the minimum message length (MML) criterion \citep[see, for instance,][]{riss,Wallace87}.
Considering the MML criterion in a Bayesian perspective and choosing Jeffreys' non-informative prior $p(\theta)$ for the mixture para\-meter,
\citet{Figueiredo02} propose minimizing the criterion\index{prior!Jeffreys}
$$
\mbox{MML}(\theta; G)= -\log p({\mathbf y}|\theta,G) -\log p(\theta|G) +\frac{1}{2} \log |I(\theta)| +\frac{\upsilon_G}{2}(1-\log(12)),
\label{chap7_4}
$$
where $I(\theta)$ is the expected Fisher information matrix which is approximated by the complete-data Fisher information matrix $I_C(\theta)$.

 As we know, for instance from Chapter~4 above,
 Jeffreys' non-informative prior does not work for mixtures. \cite{Figueiredo02}
circumvent this difficulty by only considering  the parameters of the components  whose proportion is non-zero, namely the components $g$ such that $\hat \eta_g >0$.

Assuming, for instance, that the mixture model considered arises from  the general Gaussian mixture family with free covariance matrices,
this approach leads to minimizing the criterion
\begin{align} \label{fj}
\mbox{MML}(\theta; G)= &-\log p({\mathbf y}|\theta,G) + \frac{G^\star}{2} \log \frac{n}{12} \nonumber \\
&+\frac{\mbox{dim}(\theta_g)}{2}\sum_{g:\hat \eta_g >0}\{\log(n \cdot \mbox{dim}(\theta_g)/12)+G^\star(\mbox{dim}(\theta_g)+1)\},
\end{align}
with $G^\star=\mbox{card}\{g| \hat \eta_g >0\}$.
In this Bayesian context, the approach of \citet{Figueiredo02} involves optimizing iteratively the criterion (\ref{fj}), starting from a large
number of components $G_{\max}$, and cancelling the components $g$ such that, at iteration $s$,
\begin{equation}\label{labcan}
\sum_{i=1}^n \hat \tau_{ig}^{(s)} < \frac{\mbox{dim}(\theta_g^{(s)})}{2},
\end{equation}
where  $\hat \tau_{ig}^{(s)}$ are the elements of the fuzzy classification matrix defined in (\ref{fuzzys}).
Thus, the chosen number of components   $G^\star$ is the number of components remaining at the convergence of the iterative algorithm.
This iterative algorithm could be the EM algorithm, but \citet{Figueiredo02} argue that with EM, for large $G$,
it can happen that no component has enough initial support, as the criterion for cancellation   defined in (\ref{labcan})  is fulfilled
for {\em all} $G$ components.
 Thus, they prefer to make use of the componentwise EM algorithm of \citet{CCFM},  which updates the $\eta_g$ and the $\theta_g$
sequentially: update $\eta_1$ and $\theta_1$, recompute $\tau_{i1}$ for $i=1,\ldots,n$,
update $\eta_2$ and $\theta_2$, recompute $\tau_{i2}$ for $i=1,\ldots,n$, and so on.

 \citet{Zeng14} use exactly the same approach with the completed-data  or the classification likelihood instead of the observed-data likelihood.
Thus, roughly speaking, the procedure of \citet{Figueiredo02} is expected to provide a similar number of components to the  BIC,
while the procedure of \citet{Zeng14} is expected to provide a similar number of clusters to the ICLbic presented in Section~\ref{icl}.

\subsection{Bayesian model choice based on marginal likelihoods}  \label{cfr:marlik}

\index{marginal likelihood|(}
From a Bayesian testing perspective, selecting the number of components can be interpreted as a model selection problem, given the probability
of  each  model within a collection of  all models corresponding to the different numbers of components \citep{berger:1985}.
The standard Bayesian tool for making this model choice is based on the marginal likelihood (also called {\em evidence}) of the data $\bp(\by|G)$ for
each model ${\cal M}_G$, defined in (\ref{eq:marlik}), which naturally penalizes models with more components (and more parameters)
\citep{berger:jefferys:1992}.\index{marginal likelihood}

While the BIC is often considered as one case of information criterion, 
it is important to recall (see Section~\ref{aic_bic})
 that it was first introduced by \citet{Schwartz} as an approximation to the marginal likelihood $\bp(\by|G)$.
Since this approximation does not depend on the choice of the prior $p(\thmod|G)$, it is not of direct appeal for a
Bayesian evaluation of the number of components, especially when considering
that the marginal likelihood itself can be approximated by simulation-based methods,  as
discussed in this section.

\subsubsection{Chib's method, limitations and extensions}\label{sub5:chibz}

The reference estimator for evidence approximation is Chib's
(\citeyear{chib:1995}) representation of the marginal likelihood of model ${\cal M}_G$ as\footnote{This was earlier called {\em the candidate's
formula} by Julian Besag (\citeyear{bes:can}).}
\begin{eqnarray} \label{label723}
\bp(\by|G ) = \dfrac{p(\by|\thmod^o,G)p(\thmod^o|G)}{ p (\thmod^o|\by,G)},
\end{eqnarray}
which holds for {\em any} choice of the plug-in value $\thmod^o$. While the posterior $p (\thmod^o|\by,G)$ is not available in closed form for mixtures,
a Gibbs sampling\index{Gibbs sampling}\index{Chib's evidence approximation}\index{Rao--Blackwellization}\index{candidate's formula}
decomposition allows for a Rao--Blackwellized approximation of this density \citep{robert:casella:2004} that furthermore converges at a
parametric speed, as already noticed in \citet{gelfand:smith:1990}:
$$
\hat p (\thmod^o|\by,G) = \frac{1}{\Mmc} \sum_{m=1}^{\Mmc} p (\thmod^o|\by,\zm \im{m} , G) ,
$$
where $\zm \im{m}, m=1, \ldots, M$, are the posterior draws for the latent allocations $\zm=(z_1, \ldots, z_n)$, introduced earlier in Chapter~1;
see Chapter~5 for a review of posterior sampling methods.

However, for mixtures, the convergence of this estimate is very much hindered by the fact that it requires
perfect symmetry in the Gibbs sampler, that is, complete label switching\index{label switching}  within the simulated Markov chain. When
the completed chain $(z_1\im{m} ,\ldots,z_n \im{m} )$ remains instead concentrated around one single or a subset of the
modes of the posterior distribution, the approximation of $\log \hat p (\thmod^o|\by,G) $
based on Chib's representation fails, in that it is usually off by a numerical factor of order $\text{O}(\log\,G!)$.
Furthermore, this order cannot be used as a reliable correction, as noted by \citet{neal:1999} and
\citet{fruhwirth:2006}.

A straightforward method of handling Markov chains that are not perfectly mixing (which is the usual setting) is found
in \citet{berkhof:mechelen:gelman:2003} (see also \citealp{fruhwirth:2006}, Section 5.5.5; \citealp{lee:marin:mengersen:robert:2008}) and can be interpreted as a
form of Rao--Blackwellization. The proposed correction is to estimate $\hat p (\thmod^o|\by,G)$ as an average computed over all possible
permutations of the labels, thus forcing the label switching and the exchangeability of the labels to occur in a
\lq\lq perfect\rq\rq\ manner. The new approximation can be expressed as
$$
\tilde  p (\thmod^o|\by,G)  = \frac{1}{\Mmc G!}
\sum_{\perm \in\mathfrak{S}(G)}\sum_{m=1}^\Mmc   p (\thmod^o|\by,\perm (\zm \im{m}), G) \,,
$$
where $\mathfrak{S}(G)$ traditionally denotes the set of the $G!$ permutations of $\{1,\ldots,G\}$ and where $\perm$ is
one of those permutations. Note that the above correction can also be rewritten as
\begin{equation} \label{eq_01}
\tilde  p (\thmod^o|\by,G)  = \frac{1}{\Mmc G!}
\sum_{\perm \in\mathfrak{S}(G)}\sum_{m=1}^\Mmc   p (\perm (\thmod^o) |\by,\zm \im{m}, G)  \,,
\end{equation}
as this may induce some computational savings. Further savings can be found in the im\-port\-ance sampling approach of
\citet{lee:robert:2016}, who reduce the number of permutations to be considered.\index{importance sampling}

While Chib's representation has often been advocated as a reference method for
computing the evidence, other methods abound, among them
nested sampling \citep{skilling:2007,chopin:robert:2010}, reversible jump
MCMC \citep{green:1995,richardson:green:1997}, particle filtering \citep{chopin:2002},
bridge sampling \citep{fruhwirth:2004} and path sampling\index{path sampling}\index{nested sampling}\index{bridge sampling}\index{particle filtering}
\citep{gelman:meng:1998}. Some of these methods are discussed next. 

\subsubsection{Sampling-based approximations} \label{montecar}

If $G$ is moderate, sampling-based techniques are particularly useful
for estimating the marginal likelihood of finite mixture models; see  \citet{fruhwirth:2004} and  \citet{lee:robert:2016}.
\citet{fruhwirth:2004} considered three  such estimation techniques, namely importance sampling, reciprocal importance sampling, and bridge sampling.

For sampling-based techniques, one
selects an importance density $q_G(\thmod)$ which is easy to sample from and
 provides a rough approximation to the  posterior density $p(\thmod |\ym,G)$.
 Given a suitable importance density $q_G(\thmod)$,  an importance sampling approximation to the marginal likelihood is based on  rewriting (\ref{eq:marlik}) as
$$
p(\ym | G)=\int \frac{p(\ym| \thmod, G) p(\thmod |G )}{q_G(\thmod)} q_G(\thmod) \dd \thmod.
\label{intro:model99}
$$
Based on  a sample $\thmod \im{l} \sim q_G(\thmod )$, $l=1,\ldots,L$,  from
the importance density $q_G(\thmod)$, the im\-port\-ance sampling estimator of the marginal likelihood is  given by
\begin{eqnarray}
  \hat{p}_{IS}(\ym|G) = \frac{1}{L}
\sum_{l=1}^L\frac{p(\ym|\thmod \im{l},G)p(\thmod \im{l}|G)}{q_G(\thmod \im{l})}. \label{MCE}
\end{eqnarray}
\citet{gelfand:dey:1994} introduced reciprocal importance sampling, which
is based on the observation that (\ref{label723}) can be written as
$$
\dfrac{1}{\bp(\by|G )} = \dfrac{p (\thmod |\by,G)}{ p(\by|\thmod,G) p(\thmod |G)}.
$$
Integrating both sides of this equation with respect to the importance density $q_G(\thmod)$ yields
\begin{eqnarray*} \label{label743}
\dfrac{1}{\bp(\by|G )} =\int  \dfrac{q_G(\thmod)}{ p(\by|\thmod,G) p(\thmod |G)}   p (\thmod |\by,G) .
\end{eqnarray*}
This leads to the reciprocal importance sampling estimator of the marginal likelihood, where the inverse of the ratio appearing in (\ref{MCE}) is evaluated at
the MCMC draws $\thmod \im{m}$, $m=1,\ldots,\Mmc$, and  no draws from the  importance density $q_G(\thmod)$ are required:
$$
\displaystyle  \hat{p}_{RI}(\ym|G) = \left(\frac{1}{\Mmc}
\sum_{m=1}^\Mmc \frac{q_G(\thmod \im{m})}{p(\ym|\thmod \im{m},G)p(\thmod \im{m}|G)}\right)^{-1}. \label{LRI}
$$
These two estimators are special cases of  bridge sampling \citep{meng:wong:1996}:
$$
\bp(\by|G ) = \dfrac{\Ew_{q_G(\thmod)}( \alpha(\thmod ) p(\by|\thmod,G) p(\thmod |G) )}{\Ew_{ p (\thmod |\by,G)} (\alpha(\thmod )  q_G(\thmod))},
$$
with specific functions $ \alpha(\thmod ) $. 
The (formally) optimal choice for  $ \alpha(\thmod ) $  yields the bridge sampling estimator $\hat{p}_{BS} (\ym|G)$
and combines draws  $\thmod \im{l}$, $l=1,\ldots,L$, from the importance density with MCMC draws $\thmod \im{m}$, $m=1,\ldots,\Mmc$.
Using   $\hat{p}_{IS}(\ym|G)$
as a starting value for $\hat{p}_{BS,0}(\ym|G)$, the following recursion is applied until convergence  to estimate   $\hat{p}_{BS} (\ym|G)=
  \lim_{t\rightarrow\infty}  \hat{p}_{BS, t}(\ym|G)$:
  \begin{eqnarray}
 \hat{p}_{BS, t} (\ym|G)  =
   \frac{\displaystyle  L^{-1} \sum_{l=1}^L \frac{p(\ym|\thmod \im{l},G)p(\thmod \im{l}|G) }
{L  q_G(\thmod \im{l})+M p(\ym|\thmod \im{l},G)p(\thmod \im{l}|G) /\hat{p}_{BS,t-1} (\ym|G) }}
{ \displaystyle M^{-1} \sum_{m=1}^M
\frac{q_G (\thmod \im{m} )}{L q_G(\thBF \im{m})+M p(\ym|\thmod \im{m},G)p(\thmod \im{m}|G) /\hat{p}_{BS,t-1} (\ym|G)}}.   \label{MLBSA}
\end{eqnarray}
The reliability  of these estimators depends on several factors. First, as  shown by \citet{fruhwirth:2004},
 the tail behaviour of  $q_G(\thmod)$  compared to  the mixture posterior
$p(\thmod |\ym,G)$ is relevant. Whereas the bridge sampling  estimator $\hat{p}_{BS} (\ym|G)$ is fairly robust to the tail behaviour of  $q_G(\thmod)$,
$\hat{p}_{IS}(\ym|G)$ is sensitive if   $q_G(\thmod)$  has lighter  tails than $p(\thmod |\ym,G)$,
and   $\hat{p}_{RI}(\ym|G)$ is sensitive if $q_G(\thmod)$  has fatter tails than $p(\thmod |\ym,G)$.
Second, as pointed out by \citet{lee:robert:2016},   for any of these methods  it is essential that  the importance density
$q_G(\thmod)$ exhibits the same kind of multimodality as the mixture posterior $p(\thmod |\ym,G)$ and    all modes of
the posterior density are covered by the importance density also for increasing values of $G$. Otherwise, sampling-based estimators of the marginal
likelihood  are prone to be biased for the same reason Chib's estimator is biased, as discussed in Section \ref{sub5:chibz}. 
A particularly stable estimator is obtained when bridge sampling is combined with a perfectly symmetric  importance
density $q_G(\thmod)$. 
Before the various estimators are illustrated for  three well-known data sets \citep{richardson:green:1997}, we turn to the choice of appropriate importance densities.

\paragraph*{Importance densities for mixture analysis}

 As manual tuning of the importance density  $q_G(\thmod)$  for each model under consideration is rather tedious, methods for choosing sensible importance
densities in an unsupervised  manner are needed.
 \citet{dic-etal:com}, for instance,  suggested various methods to construct  Gaussian importance densities
 from the MCMC output. However,  the multimodality of the mixture posterior
  density  with $G!$ equivalent  modes 
  evidently rules out such a simple choice.
   \citet{fru:bay}  is one of  the earliest references that  used Rao--Blackwellization to construct an
  unsupervised importance density from the MCMC output   to compute marginal likelihoods via sampling-based approaches
   and applied this idea
to model selection for linear Gaussian state space models.\index{Rao--Blackwellization}
\citet{fruhwirth:2004} extends this idea  to finite mixture and Markov switching models
 where the complete-data posterior $p(\thmod |\ym,\zm)$ is available in closed form. \citet{lee:robert:2016} discuss importance sampling
 schemes based on (nearly) perfectly symmetric  importance densities.

For a mixture distribution, where the component-specific parameters $ \theta_{g}$  can be sampled in one block
 from the complete-data posterior $p(\theta_{g}|\zm  ,\ym)$,
 Rao--Blackwellization yields the importance density
 \begin{eqnarray}
\displaystyle  q_G (\thmod)
= \frac{1}{S}  \sum_{s=1}^S  p (\eta| \zm  \im{s}) \prod_{g=1}^G   p(\theta_{g}|\zm  \im{s},\ym) ,
  \label{QMIX}
\end{eqnarray}
where $\zm  \im{s}$ are the posterior draws for the  latent allocations. 
The construction of this im\-port\-ance density is fully automatic and  it is sufficient to store the moments of these conditional densities (rather than  the
allocations $\zm$ themselves)
during MCMC sampling  for later evaluation.  This method can be extended to cases
 where    sampling  $\theta_{g}$ from  $p(\theta_{g}|\zm  ,\ym)$
 requires  two (or even more) blocks such as
 for Gaussian mixtures where $\theta_g=(\mu_g, \sigma^2_g)$ is sampled in two steps
 from $p(\mu_{g}|\sigma^2_g, \zm  ,\ym)$ and  $p(\sigma^2_g| \mu_{g}, \zm  ,\ym)$.

Concerning the number of components in (\ref{QMIX}), on the one hand   $S$  should be small for computational reasons,
because  $q_G(\thmod)$ has to be evaluated for each of the $S$ components numerous times (e.g.\
  $L$ times for the importance sampling estimator (\ref{MCE})).
On the other hand,  as mentioned above,  it is essential  that
 $q_G(\thmod)$ covers all symmetric modes of the mixture posterior, and this will require a dramatically  increasing number
of components $S$ as $G$ increases.
Hence, any of these estimators is limited to moderate values of $G$, say up to $G= 6$.

Various strategies are available to ensure multimodality in the construction of the im\-port\-ance density.
\citet{fruhwirth:2004} chooses $S=M$ and  relies on random permutation Gibbs sampling \citep{fru:mcm}\index{random!permutation}
 by  applying  a randomly selected  permutation  $\perm _m \in\mathfrak{S}(G)$
at the end of the $m$th MCMC sweep to define
a permutation  $\zm \im{s} = \perm _m (\zm \im{m})$  of the posterior draw $\zm\im{m}$ of the  allocation vector.
  The  random permutations $\perm _1, \ldots, \perm _M$   guarantee multimodality of $q_G (\thmod)$ in (\ref{QMIX}); however,
   as discussed above, it is important to ensure good mixing of the underlying permutation sampler over all $G!$ equivalent posterior  modes.
Only if $S$  is large compared to  $G!$ are all symmetric modes visited by random permutation sampling. Choosing, for instance,
 $S=S_0 G!$ ensures that each mode is visited on average $S_0$ times.


As an alternative to  random permutation sampling, approaches exploiting full permutations have been suggested; see, for example, \citet{fruhwirth:2004}.
Importance sampling schemes exploiting full permutation were discussed  in full detail in \citet{lee:robert:2016}. The definition of
a fully symmetric importance density $q_G (\thmod)$ is related to the correction for Chib's estimator discussed  earlier in (\ref{eq_01}):
\begin{eqnarray}   \label{QMIXfull}
  q_G (\thmod) = \frac{1}{S_0 G!} \sum_{\perm \in\mathfrak{S}(G)}\sum_{s=1}^{S_0}
 p (\eta| \perm ( \zm  \im{s})) \prod_{g=1}^G   p(\theta_{g}|\perm ( \zm  \im{s}),\ym) .
\end{eqnarray}
This construction, which has $S=S_0 G!$  components, is based on  a small number $S_0$ of particles $\zm  \im{s}$,
as $q_G (\thmod)$ needs to be only a rough approximation to the mixture posterior $p(\thmod |\ym,G)$ and estimators such as bridge sampling
will be robust to the tail behaviour  of $q_G (\thmod)$.  In (\ref{QMIXfull}),  all symmetric modes are visited exactly $S_0$ times.
The moments of the  $S_0$
conditional densities need to be stored  for  only one of the $G!$ permutations and, again, this construction can be extended to the case
where the components of $\theta_{g}$ are sampled in more than one block.  \citet{lee:robert:2016} discuss strategies for reducing the computational burden
associated with evaluating  $ q_G (\thmod)$.

 \citet[p.\ 146]{fruhwirth:2006} and \citet{lee:robert:2016} discuss a simplified version of (\ref{QMIXfull})  where
 the random sequence $\zm \im{s}$, $s=1,\ldots, S_0$, is substituted by   a single
optimal partition $\zm ^{\star}$ such as the maximum \textit{a posteriori} (MAP) estimator:
\begin{eqnarray*}
\displaystyle  q_G (\thmod)
= \frac{1}{G!}  \sum_{\perm \in\mathfrak{S}(G)}   p(\thmod | \perm ( \zm ^{\star}) ,\ym).
 \label{QMIXpoor}
\end{eqnarray*}
In MATLAB,  the {\sf bayesf}\index{MATLAB package!bayesf@{\sf bayesf}}  \index{bayesf@{\sf bayesf}}
package \citep{fru:bayesf} allows one to estimate
$\hat{p}_{BS} (\ym|G)$,  $\hat{p}_{IS}(\ym|G)$ and    $\hat{p}_{RI}(\ym|G)$  with  the importance density being constructed 
either as in (\ref {QMIX}) using random  permutation sampling or  as in (\ref {QMIXfull}) using full  permutation sampling.

\paragraph*{Example: Marginal likelihoods for  the data sets in \citet{richardson:green:1997}}

By way of illustration, marginal likelihoods are computed for  mixtures  of $G$ univariate normal  distributions $\Normal (\mu_g, \sigma_g^2)$ for
$G=2, \ldots, 6$  for  the acidity data, the enzyme data  and the galaxy
data   studied by \citet{richardson:green:1997} in the framework of  reversible jump
MCMC (see Section~\ref{RJMCMC} for a short description of this one-sweep method).
We use the same priors as  \citeauthor{richardson:green:1997}, namely the symmetric Dirichlet prior $\eta \sim \Dirinv{G}{1}$,
the normal prior $\mu_g\sim \Normal(m,R^2)$, the inverse gamma prior $\sigma^2_g \sim
\IG(2, C_{0} )$ and the gamma prior $C_0 \sim \Gammad (0.2, 10/R^2)$, where $m$ and $R$ are the midpoint and the
length of the observation interval. For a given $G$, full conditional Gibbs sampling is performed for $M=12{,}000$ draws
after a burn-in of 2000, by iteratively sampling from $p(\sigma^2_g|\mu_g,C_0,\zm,\ym)$, $p(\mu_g| \sigma^2_g, \zm,\ym)$,
$p(C_0|\sigma^2_1,\ldots,\sigma^2_G)$,  $p(\eta|\zm)$ and $p(\zm|\theta,\ym)$.

A fully symmetric  importance density $q_{G,F} (\theta)$ is constructed from   (\ref {QMIXfull}), where
 $S_0=100$ components are selected for each mode. For comparison, an  importance density $q_{G,R} (\theta)$
 is constructed from  (\ref {QMIX}) with
 $S= S_0 G!$, ensuring  that for random permutation sampling each mode is  visited on average $S_0$ times. However,  unlike
 $q_{G,F} (\theta)$,  the importance
 density $q_{G,R} (\theta)$  is not fully symmetric.  Ignoring the dependence between $\mu_g$  and $\sigma^2_g$,
 the component densities are  constructed from  conditionally
 independent densities, given  the $s$th draw of $(\zm,\theta_1 ,\ldots, \theta_G,C_0)$:
 $$p(\mu_g, \sigma^2_g  | \zm \im{s},\theta_g  \im{s},C_0 \im{s},\ym) =
 p(\mu_g| \sigma_g  \imarg{2}{s}, \zm \im{s},\ym)p( \sigma^2_g| \mu_g \im{s},C_0 \im{s} ,\zm \im{s},\ym) .$$
  Prior evaluation is based on the marginal prior   $p(\sigma_1^2,\ldots,\sigma_G^2)$, where  $C_0$ is integrated out.

\begin{figure}
\begin{center}
\scalebox{0.42}{\includegraphics{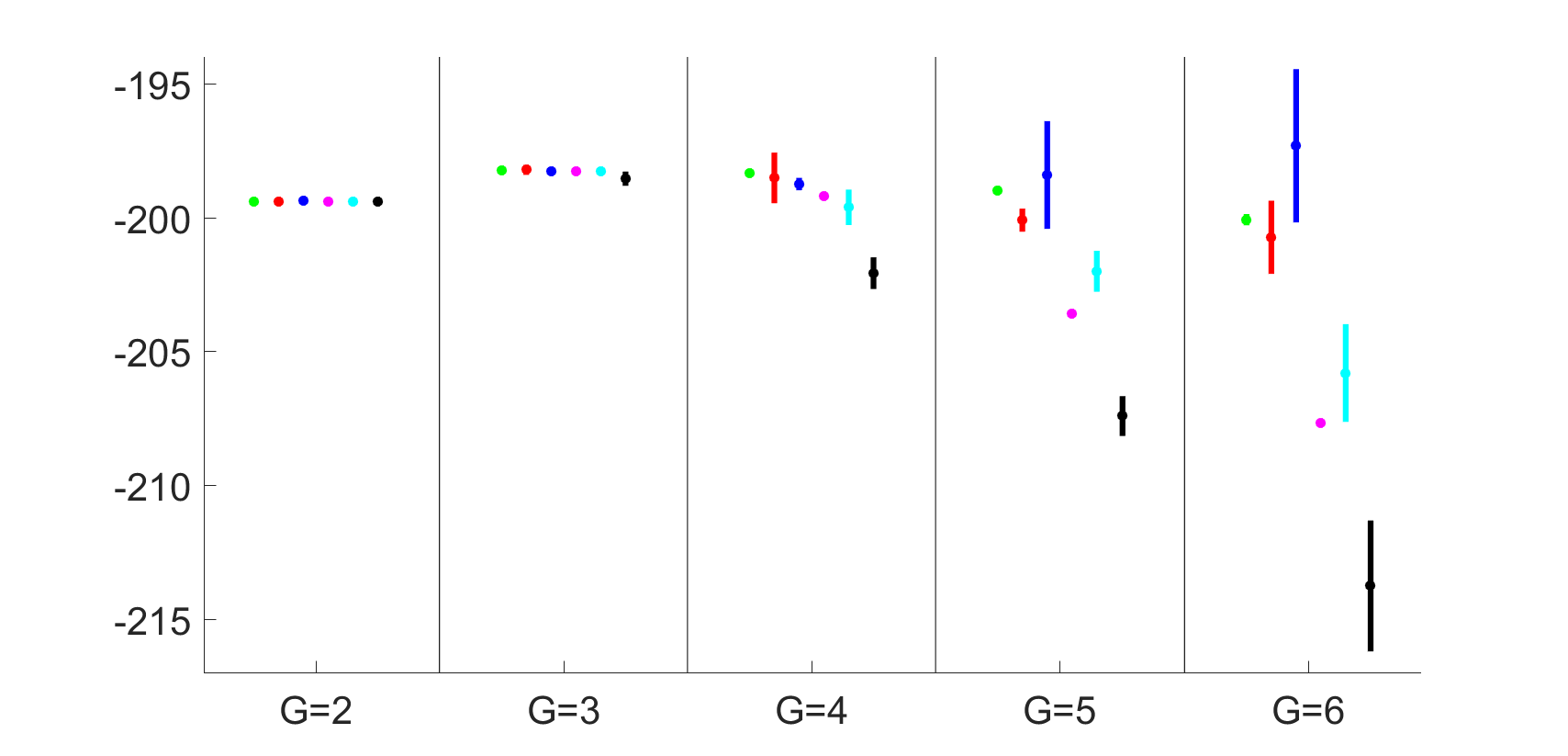}}\\
\scalebox{0.42}{\includegraphics{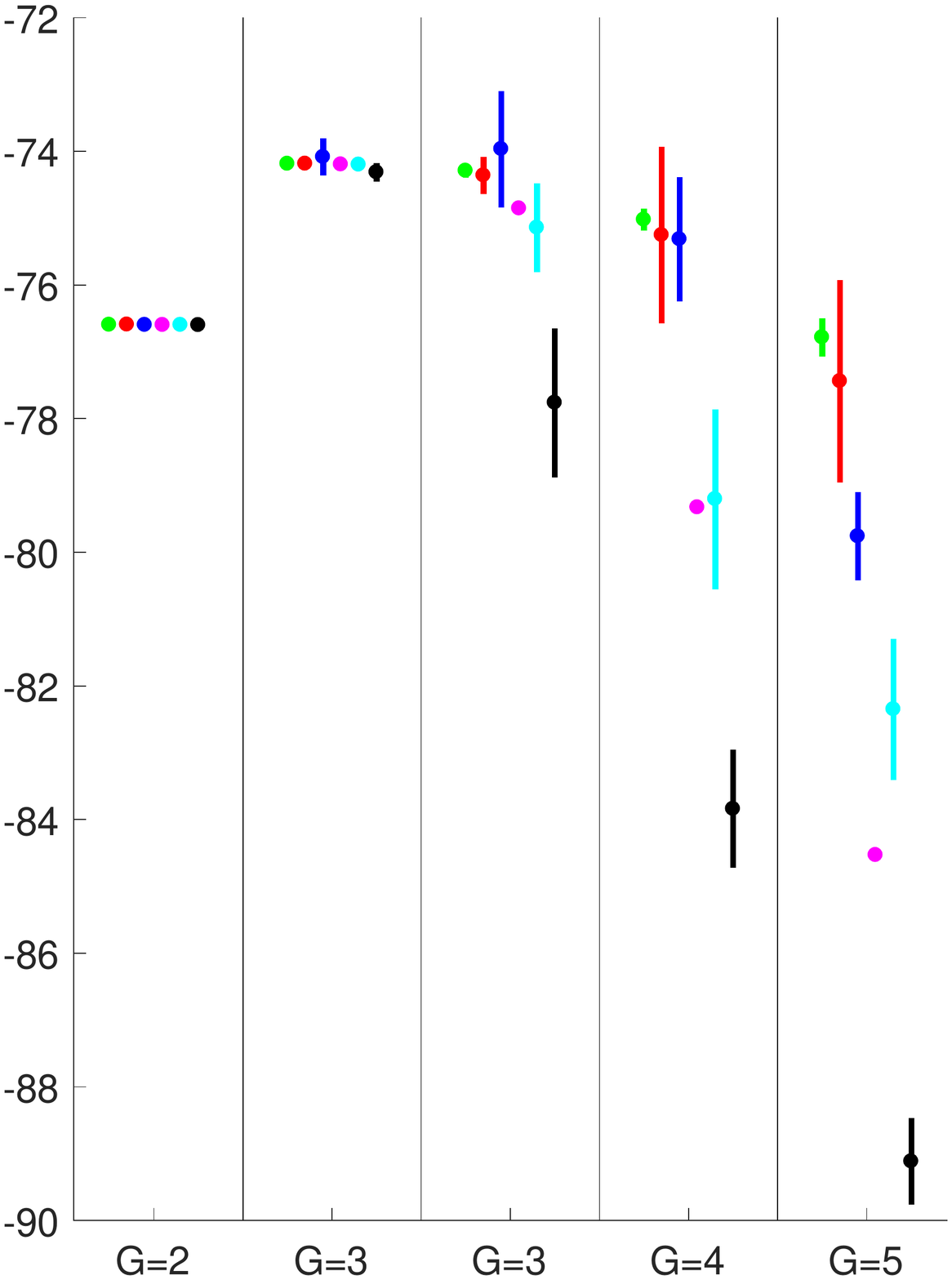}}\\
\scalebox{0.42}{\includegraphics{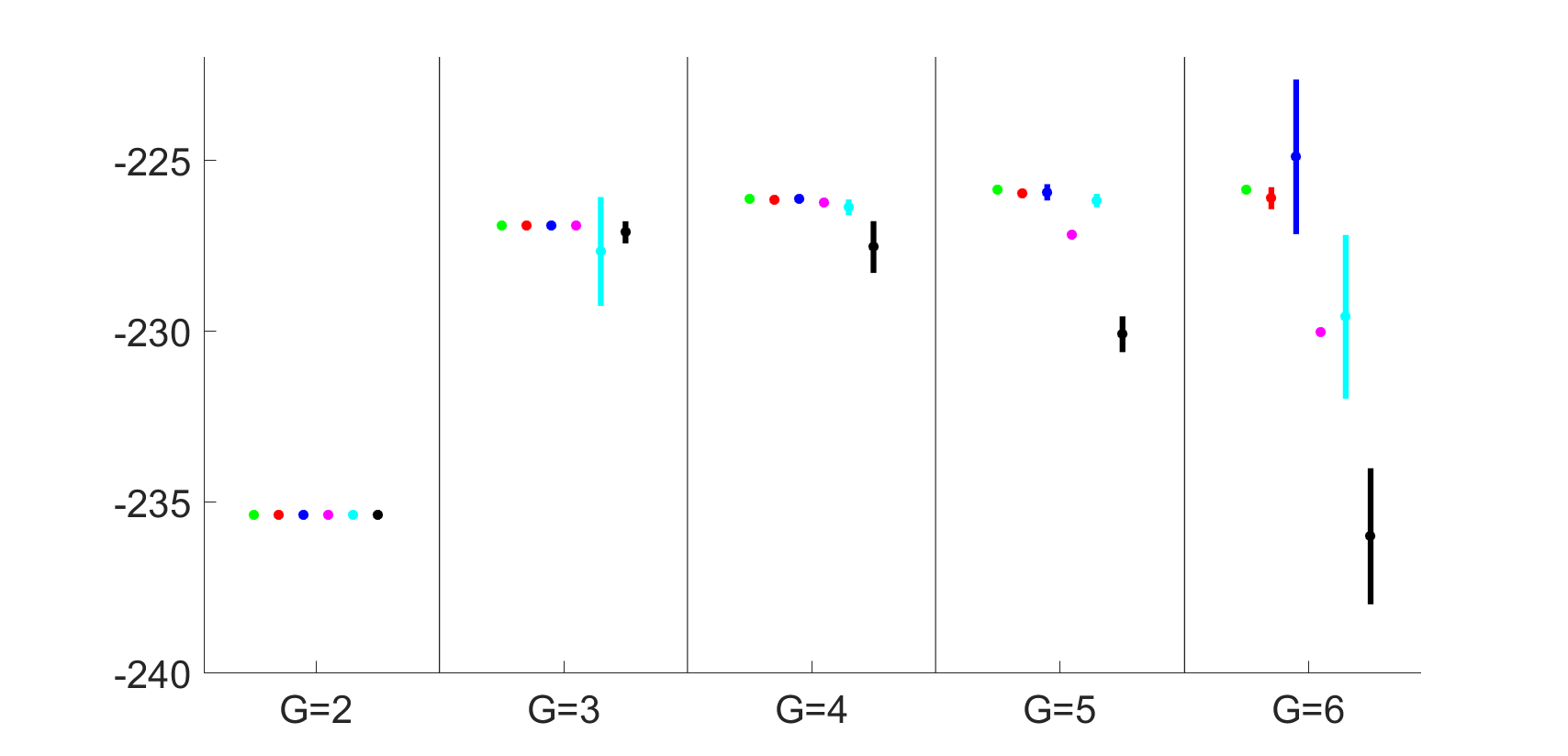}}\\
\end{center}
\caption{Marginal likelihood estimation for the benchmarks in \citet{richardson:green:1997}: the acidity data (top), the enzyme data (middle) and the galaxy
data (bottom) over $G=2,\ldots,G=6$. For each $G$,
six estimators $\log \hat{p}_{\bullet}(\ym|G) $ are given together with $\log  \hat{p}_{\bullet}(\ym|G) \pm 3 \, \mbox{SE}$ in the order
$\log  \hat{p}_{BS,F} (\ym|G)$,  $\log  \hat{p}_{IS,F}(\ym|G)$,    $\log  \hat{p}_{IS,R}(\ym|G)$, $\log  \hat{p}_{BS,R} (\ym|G)$,  $\log \hat{p}_{RI,F}(\ym|G)$
  and    $\log  \hat{p}_{RI,R}(\ym|G)$ from left to right.}\label{cfr:figml}
\end{figure}

 This yields in total six estimators,  $\hat{p}_{BS,F} (\ym|G)$,  $\hat{p}_{IS,F}(\ym|G)$ and    $\hat{p}_{RI,F}(\ym|G)$
 for full permutation sampling and  $\hat{p}_{BS,R} (\ym|G)$,  $\hat{p}_{IS,R}(\ym|G)$ and    $\hat{p}_{RI,R}(\ym|G)$ for random permutation
 sampling, for each  $G=2, \ldots, 6$.
 Results are visualized  in 
 Figure~\ref{cfr:figml},  
 by plotting  the six estimators $\log \hat{p}_{\bullet}(\ym|G) $ as well as $\log  \hat{p}_{\bullet}(\ym|G) \pm 3 \, \mbox{SE}$
 over $G$ for all three data sets.
  For each estimator, the standard errors SE are computed as in \citet{fruhwirth:2004}.
 Good estimators should be unbiased  with small standard errors and
  the  order in which the six  estimators are arranged   (which is the same for all  $G$s) is related to this quality measure.

There is a striking difference  in the reliability of the six estimators, in particular as $G$ increases.
 Reciprocal importance sampling  is particularly  unreliable
 and the estimated  values of $\log \hat{p}_{RI,R} (\ym|G)$ under $q_{G,R} (\theta)$  tend to be extremely biased  for $G\geq 4$,
 even if the  bias is reduced to a certain extent by   choosing  the fully
 symmetric importance density $q_{G,F} (\theta)$.
 Also  the two other estimators $\log \hat{p}_{IS,R} (\ym|G)$ and $\log \hat{p}_{BS,R} (\ym|G)$    tend to be biased  under $q_{G,R} (\theta)$,
and  bridge sampling  is more sensitive than  importance sampling
to choosing  an  importance density that is not fully symmetric.

Unlike for reciprocal importance sampling,  the  bias disappears for both bridge sampling  and importance sampling
 under the fully  symmetric importance density $q_{G,F} (\theta)$, and
 $\log \hat{p}_{IS,F} (\ym|G)$ and $\log \hat{p}_{BS,F} (\ym|G)$    yield  more or less identical results.
However, due to the robustness of bridge sampling with respect to the tail behaviour of
$q_{G,F} (\theta)$, we find that the standard errors of  $\log \hat{p}_{BS,F} (\ym|G)$
are often considerably smaller than the standard errors of $\log \hat{p}_{IS,F} (\ym|G)$, in particular for the enzyme data.

Based on $\log \hat{p}_{BS,F} (\ym|G)$,   marginal likelihood  evaluation  yields the following results for the three data sets.
For the acidity data,  $\log \hat{p}_{BS,F} (\ym|G=3)=-198.2 $ and   $\log \hat{p}_{BS,F} (\ym|G=4)= -198.3$   are  more less the same,
 with the log odds   of  $G=3$ over $G=4$ being equal to 0.1.
Also for the  enzyme data,   with
 $\log \hat{p}_{BS,F} (\ym|G=3)=-74.2  $ and   $\log \hat{p}_{BS,F} (\ym|G=4)= -74.3$,  the log odds   of  $G=3$ over $G=4$ are equal to 0.1.
Finally, for the galaxy data,   $\log \hat{p}_{BS,F} (\ym|G=5)= \log \hat{p}_{BS,F} (\ym|G=6)= -225.9$.
 Hence, under the prior $p(\theta|G)$ employed by  \citet{richardson:green:1997}, for all three data sets no clear distinction can be made
between  two values of $G$ based on the  marginal likelihood.  However, if the marginal likelihoods are combined with a prior on the number of components such as
$G-1 \sim \Poisson(1)$  
\citep{nob:pos}, then the log posterior odds, being equal to  1.5 for the acidity and the enzyme data and 1.8 for the galaxy data,
yield evidence for the smaller of the two values of $G$ for all three data sets.

\section{Selecting $G$ in the Framework of  Model-Based Clustering} \label{secG_clust}


Assuming that the data stem from one of the models under comparison is most
often un\-real\-istic and can be misleading when using the AIC
or BIC.  Now a common feature of standard penalized likelihood criteria is that they abstain from taking the modelling purpose into
account, except when inference is about estimating the data density. In particular, misspecification can lead to
overestimating the complexity of a model in practical situations. Taking the modelling purpose into account when selecting
a model leads to alternative model selection criteria that favor useful and parsimonious models. This viewpoint is particularly
relevant when considering a mixture model for model-based clustering; see Chapter~8 for a review of this important application of mixture models.
\index{marginal likelihood|)}

\subsection{Mixtures as partition models}   \label{sec:exp_par}

Clustering arises in a natural way when an  \iid\ sample  
is drawn from the finite mixture distribution (\ref{eq:mixOfMix1}) with weights  $\eta=(\eta_1,\ldots,\eta_G)$.
As explained in Chapter~1,  each observation $y_i$  can be associated with the component,
indexed by $z_i$, that generated this data point:
  \begin{eqnarray} \label{hiergibb1}
&z_i | \etav  \sim  \Mulnom (1, \eta_1,\ldots,\eta_G),  & \\
&  y_i|z_i \sim f_{z_i} (y_i |\theta_{z_i}). & \nonumber
\end{eqnarray}
 Let $\zm=(z_1,\ldots,z_n)$ be  the collection of all component  indicators that were used
to generate the $n$   data points $\ym=(y_1, \ldots, y_n)$.
Obviously,  $\zm$ defines a partition of the data.   A cluster\index{partition!models}
$\Cl_g =\{i: z_i = g\}$ is thus defined as a subset of the data indices $\{1,\ldots,n\}$, containing all observations
with identical allocation variables $z_i$. Hence, the indicators $\zm$ define a  partition $\parti=\{\Cl_1,\ldots,
\Cl_{\Gclust}\}$ of the $n$ data points, where  $y_i$ and $y_{j}$ belong to the same cluster if and only if $z_i=z_{j}$.
The partition $\parti$ contains $\Gclust=|\parti|$ clusters, where $|\parti|$ is the cardinality of $\parti$.
  In a  Bayesian context, \index{random!partition!models}
finite mixture models imply {\em random partitions} over the lattice
\begin{eqnarray*}
\latticez = \{ (z_1,\ldots,z_n): z_i \in \{1, \ldots, G\},i=1,\ldots,n \} \label{lattSK},
\end{eqnarray*}
as will be discussed in detail in Section~\ref{sec_clust}.

In model-based clustering,  a finite mixture model is applied  to recover the  (latent) allocation indicators  $\zm$
from  the data and to estimate  a suitable   partition of the data.
A useful quantity in this respect is  the  so-called fuzzy classification matrix $\tau$.
The elements $\tau_{ig}$, with $i=1,\ldots,n$ and $g=1,\ldots, G$,  of  $\tau$  are equal to  the conditional probability
that observation $y_{i}$ arises from
component $g$ in a mixture model of order $G$ given $y_i$:
\begin{equation} \label{tau}
\tau_{ig}=\Prob(z_{i}=g|y_i, \theta) = \Prob(z_{ig}=1|y_i, \theta)=  \frac{\eta_{g}f_g(y_{i} \mid {  \theta}_{g})}
 {\sum_{j=1}^{G}\eta_{j}f_j(y_{i} \mid {  \theta}_{j})},
\end{equation}
where $z_{ig}=\indic{z_i=g}$.
The entropy  $\ENT(\thmod;  G)$  \index{entropy}
 corresponding to  a  fuzzy classification matrix $\tau$ 
 is defined as
\begin{equation} \label{ent}
\ENT(\thmod;  G)=-\sum_{g=1}^{G}\sum_{i=1}^{n}\tau_{ig} \log \tau_{ig} \geq 0 .
\end{equation}
Both $\tau$ and  $\ENT(\thmod;  G)$ are data-driven measures of  the ability of  a  $G$-component mixture  model to provide a relevant partition of the data.
 If the mixture components are well separated for a given $\thmod$, then the classification matrix $\tau$ tends to define a  clear partition of
 the data set  $\ym=(y_{1},\ldots,y_{n})$, with  $\tau_{ig}$ being close to 1  for one component and close to 0 for all other components.
 In this case,  $\ENT(\thmod; G) $ is close to 0.
 On the other hand, if the mixture components are poorly separated, then
 $\ENT(\thmod;G)$ takes values larger than zero.  The maximum value
  $\ENT(\thmod;G)$ can take is $n \log G$, which is the entropy of  the uniform  distribution
 which assigns  $y_i$ to all $G$ clusters with the same probability  $\tau_{ig}\equiv 1/G$.

In a Bayesian context,  the fuzzy classification matrix is instrumental for joint estimation of  the parameter
$\theta$ and $\zm$ within Gibbs sampling using data augmentation \citep[see, for example,][]{robert:casella:2004}.
In a frequentist framework, the estimated classification matrix $\hat{\tau}$,
 given a suitable estimate  $\hat{\theta}_G$ of the mixture parameters $\theta$
 (e.g.\ the MLE), can be used to derive an estimator $\hat{\zm}$  of the
partition of the data; see also Chapter~8. 
As will be discussed in Section~\ref{sec:class-IC}, the  entropy of the estimated classification matrix $\hat{\tau}$
plays an important role in defining information criteria for choosing $G$ in a clustering context.

\subsection{Classification-based information criteria}  \label{sec:class-IC}

As discussed in Section~\ref{aic_bic}  within the framework of  density estimation, the BIC enjoys several desirable properties;
 however,  within cluster analysis it shows a tendency to overestimate $G$; see, for instance, \citet{Celeux96}.
  The BIC does not take the clustering purposes for assessing $G$ into account,   regardless of the separation of the clusters.
To overcome this limitation, an attractive possibility is to select $G$ so that the resulting mixture model leads to the
clustering of the data with the largest evidence. This is the purpose of
various classification-based information criteria such as  the  integrated complete-data likelihood criterion that are discussed in this subsection.


In a classification context, it   is useful  to state a simple relation linking
the log of the observed-data density  $p(\ym|\thmod)$ and the complete-data   density  $p(\ym,\zm|\thmod)$.\index{likelihood!complete-data}
The observed-data log likelihood  of $\thmod$ for a sample $\ym$,  denoted by $\lliko (\thmod;G)$,  is given by
\begin{equation*}
\lliko (\thmod;G) =\sum_{i=1}^{n} \log \left[\sum_{g=1}^{G} \eta_{g}f_g(y_{i}
\mid {\mathbf \theta}_{g})\right],
\end{equation*}
whereas the  complete-data  log likelihood  of $\thmod$ for the complete sample $(\ym,\zm)$, denoted by $\llikc (\thmod;G)$,  reads
\begin{equation*}
\likcom =\sum_{i=1}^{n}  \sum_{g=1}^{G} z_{ig}  \log( \eta_{g}f_g(y_{i}
\mid {\mathbf \theta}_{g}) ),
\end{equation*}
where $z_{ig} = \indic{z_i=g}$, $g=1,\ldots, G$.  These log likelihoods are linked in  the following way:
\begin{equation} \label{hata}
\likcom = \lliko (\thmod;G) -  \ECz,
\end{equation}
where
\begin{eqnarray*}
 \ECz
 =-\sum_{g=1}^{G}\sum_{i=1}^{n}z_{ig} \log \tau_{ig} \geq 0.
\end{eqnarray*}
Since $\Ew (z_{ig}|\thmod, y_i) = \Prob(z_{ig}=1|\thmod, y_i)=\tau_{ig}$,  we obtain that the expectation of $ \ECz$  with respect to
 the conditional distribution $p(\zm|\ym,\thmod)$ for a given $\thmod$ is equal the entropy
$\ENT(\thmod;  G)$  defined in (\ref{ent}).  Hence, the  entropy can be regarded as  a penalty for the observed-data likelihood in cases where the
resulting clusters are not well separated.

\subsubsection{The integrated complete-data likelihood criterion} \label{icl}

The integrated (complete-data) likelihood related to the complete data $(\ym,
\zm)$ is\index{ICL}\index{integrated!completed likelihood (ICL)}
\[
p(\ym, \zm  \mid G)=\int_{\Theta_{G}}p (\ym, \zm \mid G,\thmod)
p(\thmod \mid G) \dd \thmod,
\]
where
\[
p (\ym, \zm \mid G, \thmod )= \prod_{i=1}^n p(y_i, z_i \mid G,\thmod)
=  \prod_{i=1}^n \prod_{g=1}^G \eta_g^{z_{ig}}\left[f_g(y_i \mid {\mathbf
\theta}_g)\right]^{z_{ig}}.
\]
This integrated complete-data likelihood (ICL)
takes the missing data $\zm$ into account and can be expected to be relevant for choosing $G$ in a clustering context.
However,  computing the ICL is challenging  for various reasons.
First, computing the ICL involves an integration in high dimensions. Second,
 the labels $\zm$ are unobserved (missing) data.
\newcommand{\thetastar}{\hat \thmod^{\zm}}
 To approximate the ICL, a BIC-like approximation is possible \citep{BCG00}:
\[
\log p(\ym, \zm \mid G)\approx
\log p (\ym, \zm \mid G,\thetastar) - \frac{\upsilon_{G}}{2} \log n,
\]
where
\[
\thetastar =\arg \max_{\thmod} p (\ym, \zm  \mid G,\thmod),
\]
and $\upsilon_G$ is  the number of free parameters of the mixture model ${\cal M}_G$.
Note that  this approximation involves the complete-data likelihood,  $\liklik=p (\ym, \zm \mid G, \thmod )$;
however,  $\zm$ and, consequently,  $\thetastar$ are unknown.
First,  approximating
$\thetastar \approx  \hat \thmod_G$, with $\hat \thmod_G$ being the MLE  of the $G$-component mixture parameter $\thmod$,
is expected to be valid for well-separated components.
Second, given  $ \hat \thmod_G$,   the missing data $\zm$  are imputed using  the MAP estimator
$\hat \zm=\mbox{MAP}(\hat \thmod_G)$ defined by
\[
\hat z_{ig}= \left\{ \begin{array}{ll} 1, & \mbox{if } \mbox{argmax}_{l}  \tau_{il} (\hat \thmod_G)=g,\\ 0 ,& \mbox{otherwise.} \end{array} \right.
\]
This  leads to the criterion
\[
\mbox{ICLbic}(G)=  \log p (\ym,\hat \zm  \mid G,\hat \thmod_G )- \frac{\upsilon_{G}}{2} \log n.
\]
Exploiting (\ref{hata}), one obtains  that
the   $\mbox{ICLbic}$ criterion takes the form of a BIC criterion,  penalized by the estimated entropy
\[
\ENT (\hat \thmod_G;G)=-\sum_{g=1}^{G}\sum_{i=1}^{n}\hat \tau_{ig} \log \hat \tau_{ig} \geq 0,
\]
with $\hat \tau_{ig}$ denoting the conditional probability that $y_{i}$ arises from the $g$th mixture component
$(i= 1, \ldots, n$, $g=1, \ldots, G)$ under the parameter $\hat \thmod_G $; see (\ref{tau}).

Because of this additional entropy term, the $\mbox{ICLbic}$ criterion favours values of $G$ giving rise to partitions of  the data with
the highest evidence. In practice, the $\mbox{ICLbic}$ appears to provide a stable and reliable
estimation of $G$ for real data sets and also for simulated data sets from mixtures when the components do not
overlap too much.  However, it should be noted that  the $\mbox{ICLbic}$, which is not concerned with discovering the true number of mixture components, can
underestimate the number of components for simulated data arising from mixtures with poorly separated components.

\paragraph*{An illustrative comparison of the BIC and ICLbic}
Obviously, in many situations where the mixture components are well separated, the BIC and ICLbic select the same number of mixture components.
But the following small numerical example aims to illustrate a situation where these two criteria give different answers.

We start from a benchmark (genuine) data set known as the {\em Old Faithful Geyser}.  Each of the 272 observations
consists of two measurements: the duration of the eruption and the waiting time before the next eruption of the Old
Faithful Geyser, in Yellowstone National Park, USA.  We consider a bivariate  Gaussian mixture model with component densities $\Normal (\mu_k,\Sigma_k)$ with unconstrained
covariance matrices $\Sigma_k$.

\begin{figure}[t!]
\centering
\includegraphics[width=5.5cm]{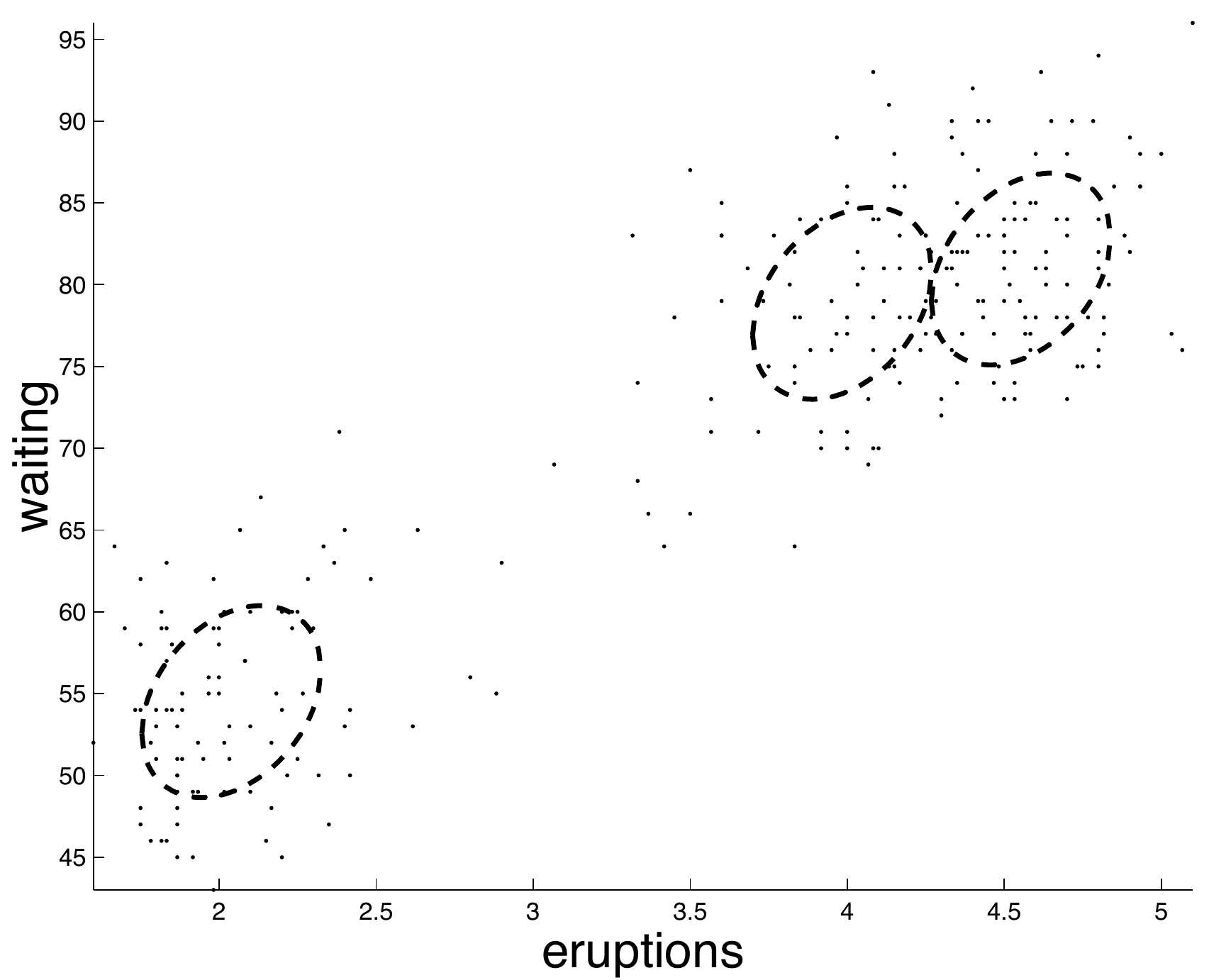}	
\hspace{1.0cm}
\includegraphics[width=5.5cm]{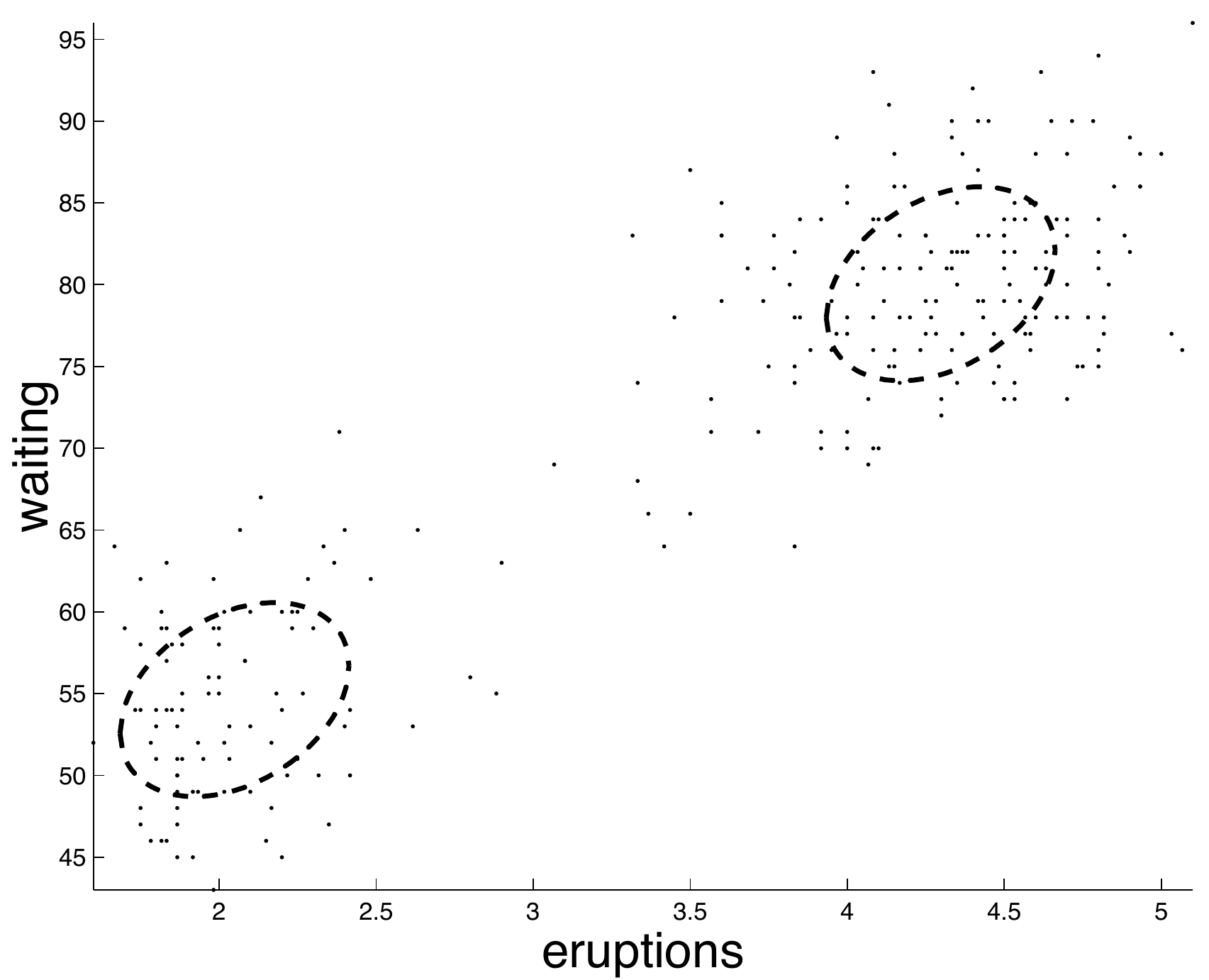}	
\caption{Cluster ellipses for the Old Faithful Geyser data:  (left) the BIC solution; (right) the ICLbic solution.} \label{geyser}
\end{figure}

For this data set, Figure \ref{geyser} shows that the ICLbic selects with a large evidence $G=2$, while the BIC slightly prefers
$G=3$ to $G=2$.  The BIC solution with $G=3$ components appears to model deviations from normality in one of the two obvious
clusters, rather than a relevant additional cluster.

\subsubsection{The conditional classification likelihood}

In a model-based clustering context where a cluster is associated with a mixture component,\index{likelihood!conditional classification}
it is sensible in view of (\ref{hata}) to maximize the conditional expectation of the
complete-data log likelihood \citep{Baudry15}, 
 \begin{equation*}
\log \lcc (\theta; G)=\Ew_{\zm} ( \likcom )  =  \lliko(\thmod; G) -\ENT(\thmod; G),
\end{equation*}
rather than the observed-data log likelihood function $ \lliko (\thmod; G)$. This can be done
through an EM-type algorithm where the M step at iteration $s+1$ involves finding
\begin{equation}  \label{fuzzys}
\theta^{(s+1)}\in\argmax_{\thmod \in\Theta_G} \left( \lliko(\thmod; G) + \sum_{i=1}^n\sum_{g=1}^G  \tau_{ig}^{(s)}\log\tau_{ig} \right) ,
\end{equation}
where the $\tau_{ig}$ are defined as in (\ref{tau})
and
\[
\tau_{ig}^{(s)} =  \frac{\eta_{g}^{(s)}f_g({  y}_{i} \mid {\mathbf \theta}_{g}^{(s)})} {\sum_{j=1}^{G}\eta_{j}^{(s)}f_j({
y}_{i} \mid {\mathbf \theta}_{j}^{(s)})}.
\]
This M step can be performed by using an adaptation of the so-called Bayesian expectation maximization (BEM)
of \citet{Lange99}.  The resulting algorithm inherits the fundamental property of EM to increase the criterion
$\log \lcc (\theta)$, which does not depend  on $\zm$, at each iteration.

In this context, \citet{Baudry15} considered choosing $G$ from a penalized criterion of the form
\[
\lcc\text{-ICL}(G)=  -  \log \lcc (\MLccE\thmod_G ; G)  + \pen(G) ,
\]
where 
$\MLccE\thmod_G= \arg \max_{\thmod}\log \lcc (\thmod ; G)$.
Under standard regularity conditions and assuming that  $ \pen:\{1,\ldots,\Gmax\}\rightarrow\mathbb{R}^+$ satisfies
\begin{equation*}
\begin{cases}
\pen(G)=\pop(n), \quad\text{as } n\rightarrow\infty ,\\
\bigl(\pen(G)-\pen(G')\bigr) \xrightarrow[n\rightarrow \infty]{\mathbb{P}} \infty,\quad\text{if } G'<G,
\end{cases}
\end{equation*}
\citet{Baudry15} proved that
\begin{equation*}
\mathbb{P}\bigl[ \widehat G \neq G_0 \bigr] \xrightarrow[n\rightarrow \infty]{} 0,
\end{equation*}
where $\hat G =\min\argmin_G \lcc\text{-ICL}(G)$ and $G_0$ is the minimum number of components such that the bias of the models is stationary for $G \geq G_0$,
\begin{equation*}
G_0 = \min\argmax_{G} \Ew_{p_0}\bigl[ {\llikc}(\theta_G^0)  \bigr],
\end{equation*}
with
\begin{equation*}
  \theta_G^0 = \argmin_{\theta\in\Theta_G} \Bigl\{ d_\text{KL}\bigl(p_0,p(\p \thmod )\bigr) + E_{p_0}\bigl[\EC(\thmod; G)\bigr] \Bigr\},
\end{equation*}
$d_\text{KL}\bigl(p_0,p(\p \thmod)\bigr)$ being the Kullback--Leibler distance between the true distribution $p_0$ of the data
and the mixture distribution with parameter $\thmod$.
Moreover, \citet{Baudry15} deduces that, by analogy with the BIC, an interesting identification criterion to be minimized is 
\begin{equation*}
\lcc\text{-ICL}(G) =   -\log \lcc (\MLccE\thmod_G; G)  + \frac{\upsilon_G}{2}\log n.
\end{equation*}
The criterion ICLbic can thus be viewed as an approximation of $\lcc$-ICL. Therefore, the criterion $\lcc$-ICL underlies a
notion of class that is a compromise between the \lq\lq mixture component\rq\rq\ and the \lq\lq cluster\rq\rq\ points of view.

\subsubsection{Exact derivation of the ICL}  \label{exaICL}

Like the BIC, the ICL has been defined in a Bayesian framework, but its asymptotic approximations ICLbic and $\lcc$-ICL
are not intrinsically Bayesian,  since they do not depend on the associated prior distribution. However, if the mixture components
belong to the exponential family, it is possible to get closed-form expressions for the ICL (see \citealp{BCG10}, or \citealp{BFR15}).
With such closed-form expressions, 
it is possible to compute the ICL values by replacing the missing labels
${\mathbf z}$ with their most probable values using the MAP operator after estimating the parameter $\hat{\thmod}_G$ as the
posterior mode or the MLE \citep[see][]{BCG10}. An alternative is to optimize the exact ICL in $\zm$.
The limitations of  approaches based on exact ICL computing are  twofold.

\paragraph{Choosing non-informative prior distributions}

\index{prior!non-informative}
Except for categorical data which involve mixtures of multivariate discrete distributions, there is no proper consensual
non-informative prior distribution for other classes of mixture models such as
Gaussian or Poisson mixture models (see Chapter~4). It is obviously possible to
choose exchangeable weakly informative hyperparameters
with conjugate prior distributions for the parameters of the mixture components.
However, the posterior distribution and thus the resulting ICL values will inevitably
depend on these hyperparameters.  For the latent class model on categorical data, deriving the exact ICL is
easier, since the non-informative   conjugate Dirichlet prior  distributions   $\Dirinv{G}{\ed{0}}$    
are proper for the weight distribution of the mixture.
Following the recommendation of \citet{SFR11}, it has been demonstrated that choosing $e_0=4$ is expected to provide  a stable selection of $G$
\citep[see, for instance,][]{KBCG15}.  Numerical experiments on simulated data proved that exact ICL computed with plug-in
estimates  $\hat{\thmod}_G$  of the parameter could provide different and more reliable estimation of $G$ than the ICLbic for small
sample sizes.  Thus, when conjugate non-informative prior distributions are available, deriving a non-asymptotic
approximation of ICL can be feasible.

\paragraph{Optimizing the exact ICL}

Several authors have considered the direct optimization of the exact ICL in $\zm$ without
estimating $\thmod$. \citet{BFR15}, \citet{Come15} and \citet{WFL15} have proposed greedy algorithms, while \citet{TSBCG06}
proposed using evolutionary optimization algorithms. At this point, it is important to remark that the optimization problem has to
be solved in a search space with about $O(\Gmax^n)$ elements, where $\Gmax$ is the maximum number of components allowed.
This means that the optimization problem becomes quite formidable for $n$ large. In addition, the proposed greedy algorithms
are highly sensitive to the numerous local optima and have only been experimented with for moderate sample sizes. This is the
reason why evolutionary algorithms are expected to be useful but they need to be calibrated (to choose the tuning
parameters) and are expensive in computing time.

\subsection{Bayesian clustering} \label{sec_clust}


In the context of Bayesian clustering \citep[see][for an excellent review]{lau-gre:bay},
where the allocation indicator $\zm=(z_1, \ldots, z_n)$ is regarded as a latent variable,
 a finite mixture model implies {\em random partitions} over the lattice $\latticez$.
Hence, for a given order $G$ of the mixture distribution (\ref{eq:mixOfMix1}),   both the prior density  $p(\zm|G)$ and the  posterior density $p(\zm|G, \ym)$ 
are   discrete distributions   over the lattice $\latticez$.
Although this induces a change of prior modelling, \citet{lau-gre:bay} discuss Bayesian nonparametric (BNP; see Chapter 6)
methods to estimate the number of
clusters\index{number of clusters}. 
We discuss  the BNP perspective further in Section~\ref{sec:BaNo} and refer to Chapter~6 for a comprehensive treatment.

For a finite mixture model,  the  Dirichlet prior $\etav \sim \Dir (\ed{1}, \ldots, \ed{G})$ on the weight distribution strongly determines
what the prior distribution $ p(\zm|G)$ looks like. To preserve symmetry with respect to relabelling, typically the symmetric Dirichlet prior  $\Dirinv{G}{\ed{0}}$
 is employed, where $\ed{1}= \ldots= \ed{G}=\ed{0}$.
The  corresponding prior $ p(\zm|G) = \int \prod_{i=1}^n p(z_i| \etav) \dd \, \etav$   is given by 
\begin{eqnarray}
\displaystyle p( \zm| G) = \frac{\Gamfun{G \ed{0}}}{\Gamfun{n+G \ed{0}}\Gamfun{\ed{0}}^{\Gclust}} \prod_{g: n_g>0} \Gamfun{n_g+\ed{0}} ,
\label{intro:dirich:ml}
\end{eqnarray}
where $n_g = \sum_{i=1}^n \indic{z_i=g}$ is the number of observations in cluster $g$ and $\Gclust$ is defined
as the number of non-empty clusters,
\begin{eqnarray} \label{Gplus}
\Gclust = G - \sum_{g=1}^G  \indic{n_g=0}.
\end{eqnarray}
As mentioned earlier, in model-based clustering
interest lies in estimating  the number of clusters  $\Kn$
in the $n$  data points  rather than the number of    components   $G$
 of the mixture distribution (\ref{eq:mixOfMix1}), and it is  important  to distinguish between both quantities.
 Only a few papers  make this  clear distinction between the number of
 mixture components $G$ and the number of  data cluster $\Gclust$ for finite mixture models
\citep{nob:pos,mal-etal:ide,mil-har:mix,fru-mal:fro}.

A common criticism concerning the application of  finite mixture models in a clustering context is that the
number of  components $G$ needs to be known \textit{a~priori}.
%
However, what is yet not commonly understood is  (a)  that  the really relevant question is whether or not the {\em number of clusters}  $\Gclust$ in the data
is known \textit{a~priori}  and (b)  that  even a finite mixture with a fixed value of $G$ can imply a random prior  distribution
on $\Gclust$.
 By way of further illustration, let  $n_g =\sum_{i=1}^n \indic{z_i=g}$   be the number of observations generated by the  components  $g=1,\ldots,G$.
Then  (\ref{hiergibb1}) implies that  $n_1,\ldots,n_G$ follow a multinomial distribution:
 \begin{eqnarray} \label{knmul}
n_1  ,\ldots, n_{G}  | \etav  \sim  \Mulnom (n, \eta_1,\ldots,\eta_G).
\end{eqnarray}
Depending on the weights  $\etav=(\eta_1,\ldots,\eta_G)$ appearing in the mixture distribution (\ref{eq:mixOfMix1}),
  multinomial sampling according to (\ref{knmul})  may lead to partitions with $n_g$ being zero, leading to so-called   \lq\lq empty components\rq\rq .
   In this case,   fewer than $G$  mixture components were used to generate the $n$ data points which
contain  $\Gclust$  non-empty clusters, where  $\Gclust$ is defined as in  (\ref{Gplus}).

In a Bayesian framework towards finite mixture modelling,  the Dirichlet prior $\etav \sim \Dirinv{G}{\ed{0}}$ on the
 component weights controls whether, {\em a priori},  $\Gclust$ is equal to $G$ and no empty components occur.
In particular, if  $e_0$ is close to 0,  then $\Gclust$ is a random variable taking   \textit{a~priori} values smaller than $G$ with
 high probability.
Exploiting the difference between   $\Gclust$ and  $G$  in an overfitting mixture with a prior on the weight distribution
 that strongly shrinks redundant component weights towards 0
 is a cornerstone of the concept of sparse finite mixtures\index{mixtures!sparse} \citep{MWFSG16} which will be discussed in Section~\ref{sec:sfm} as
a  one-sweep method to determine  $\Gclust$ for a fixed  $G$.

 \begin{figure}[t!]
\begin{center}
\scalebox{0.7}{\includegraphics{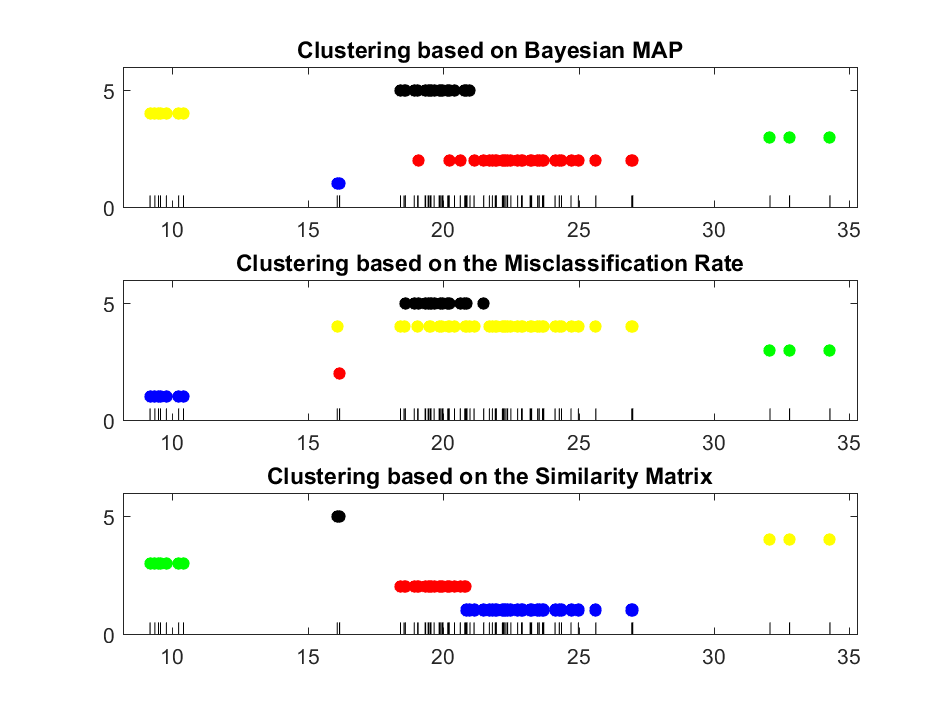}}
\end{center}
\caption{Bayesian clustering of the  galaxy data \citep{roe:den}, assuming a Gaussian mixture with $G=5$ components. The data are indicated through a rug plot.
Partitions resulting from the MAP estimator (top),
the minimum risk estimator (middle) and   minimizing Binder's loss function.}\label{cfr:figclass}
\end{figure}

In Bayesian clustering (rather than Bayesian mixture estimation), the main object of interest is the
(marginal) posterior of the allocations $\zm$, 
rather than the (marginal) posterior distribution of the mixture
parameters $\thmod$. 
Depending on the mixture under investigation, the integrated
likelihood\index{integrated!likelihood}\index{likelihood!integrated} $ p(\ym |\zm,G)$ for $G$ known may be available in closed
form, in particular, if  the component densities $f_g(y|\theta_g)$  come from exponential families and a conditionally
conjugate prior $p(\theta_g)$ is employed  for $\theta_g$.
As noted, for instance,  by \citet{cas-etal:mix}, for many mixture models it is then possible to derive an
explicit form for the marginal posterior $p(\zm| \ym, G)$ of the indicators $\zm$, where dependence on the
parameter $\thmod$ is integrated out. By Bayes' theorem, the marginal posterior $p(\zm|\ym,G)$ is given by
\begin{eqnarray}
& \displaystyle p(\zm|\ym,G) 
 \propto p(\ym |\zm,G) p(\zm|G),& \label{margSiv}
\end{eqnarray}
where  the integrated prior $ p(\zm |G)$ is given by (\ref{intro:dirich:ml}) and  the integrated likelihood $p(\ym |\zm, G)$ takes the form
\begin{eqnarray}
p(\ym |\zm,G) = \int p(\ym|\zm,\theta_1, \ldots, \theta_G,\eta, G) p(\theta_1,
\ldots, \theta_G ,\eta |G ) \dd (\theta_1, \ldots, \theta_G ,\eta) .  \label{margyz}
\end{eqnarray}
To explore the posterior  of the allocations,  efficient methods to sample from the posterior $p(\zm|\ym,G)$ are needed, and some of these
methods  will be discussed in
Section~\ref{sec:allocation}.
This exploration is quite a computational challenge, as the size of the  lattice $\latticez$ 
increases rapidly with both the number $n$ of observations and the number $G$
of components and is given by the Bell number.  For $n=10$ and $G=3$, for
instance, there are 59,049 different allocations $\zm$, whereas for $n=100$ and
$G=3$ the number of different allocations is of the order of $5\cdot10^{47}$.
This means that it is impossible to visit all possible partitions  $\parti$
during posterior sampling and many partitions are visited at best once.

This large set of partitions raises the question of how to summarize  the posterior  $p(\zm|\ym, G)$, given posterior simulations.
Common summaries are based on  deriving point estimators $\hat{\zm}$, such as the MAP estimator,
the minimum risk estimator or the partition minimizing Binder's loss function \citep{bin:bay},
see Chapter~8, for more  details. However, these estimators (even if they differ)
 do not fully reflect  the uncertainty in assigning observations to clusters.

By way of illustration, a mixture of univariate Gaussian distributions is used for
Bayesian clustering of the galaxy data \citep{roe:den}, assuming that $G=5$ is
fixed. Prior specification follows  \citet{richardson:green:1997}, and 12,000
draws from $p(\zm|\ym,G)$ are obtained using full conditional Gibbs sampling.
In Figure~\ref{cfr:figclass},   various point estimators $\hat{\zm}$  derived
from the posterior draws of $\zm$ are displayed, together with a rug plot of
the data.    While the MAP estimator   and   the estimator
minimizing Binder's loss function are invariant to label switching, the minimum
risk estimator is based on an identified model. Label switching is  resolved by
applying $k$-means clustering to the point process representation of the MCMC
draws of $(\mu_g, \sigma_g)$.   Classification over the various point
estimators $\hat{\zm}$ is stable for observations in the two clusters capturing
the tails,
but the classification  for  observations in  the centre of the distribution tends to be  rather different.

To quantify such uncertainty, \cite{wad-gha:bay} develop not only appropriate
point  estimates, but also credible sets  to summarize the posterior
distribution of the partitions based on decision- and information-theoretic
techniques.

\subsection{Selecting $G$ under model misspecification}

Mixture models\index{model!misspecification}\index{misspecification} are a very
popular tool for model-based clustering, in both the frequentist and Bayesian
frameworks. However, success in  identifying meaningful clusters in the
data very much hinges on  specifying sensible component densities, and Bayesian
inferences towards estimating the number of clusters are sensitive to
misspecifications of the component densities, as are most penalized
likelihood criteria discussed in the previous subsections. Most commonly, a
finite mixture model with (multivariate) Gaussian component densities is fitted
to the data to identify homogeneous data clusters within a heterogeneous population:
\begin{align}
  y \sim  \sum \limits_{g=1}^G \eta_g \Normal  (\mu_g,\Sigma_g) . \label{eq:mixnormmul}
\end{align}
Similarly to the likelihood approach, Bayesian cluster analysis has to address
several  issues.  First, as discussed above, even if we fit  a  correctly
specified  mixture model   (\ref{eq:mixOfMix1})  to data generated by this
model, an estimate of the number of components $G$ will not necessarily be a
good estimator of  the number of clusters $\Gclust$ in the data, and  a more
reliable estimate is obtained when exploring the partitions. 

However,  problems with the interpretation of $\Gclust$ might nevertheless occur, in particular if the
component density is misspecified and several components have to be merged to address this misspecification.
A typical example is fitting the   multivariate Gaussian mixture distribution  (\ref{eq:mixnormmul})
to data such as the {\em Old Faithful Geyser} data. As shown in Figure~\ref{geyser},
 more than one Gaussian component is needed  to capture departure from normality such as skewness and excess
kurtosis for one of the two clusters.
As discussed before, the BIC is particularly sensitive to this kind of misspecification, and classification-based information criteria
  such as the ICL criterion introduced  in Section~\ref{icl}  are more robust in this respect.

In both Bayesian and frequentist frameworks,  misspecification  has been
resolved by  choosing more flexible distributions for the components densities.
Many papers demonstrate the usefulness of mixtures of parametric non-Gaussian
component densities  in this context (see \citealp{fru-pyn:bay}, and
\citealp{lee-mcl:emm}, among many others),   and  Chapter~10 also addresses this problem.
Unsurprisingly, the estimated $\Gclust$ of   such a non-Gaussian mixture often
provides a much better  estimator of the number of clusters than does the
Gaussian mixture.   With respect to inference, the Bayesian framework offers a
slight advantage, as MCMC methods are able to deal with non-standard component
densities in a more flexible way than the EM algorithm.

In higher dimensions it might be difficult to choose an appropriate parametric
distribution for characterizing a data cluster, and  mixture models with more
flexible (not necessarily parametric) cluster densities turn out to be useful.
The  mixture of Gaussian mixtures  approach, for instance,  exploits the
ability of normal mixtures to accurately approximate a wide class of
probability distributions,  and  models the non-Gaussian cluster distributions
themselves by Gaussian mixtures.  This introduces a hierarchical framework
where in the upper level a non-Gaussian mixture is fitted as in
(\ref{eq:mixOfMix1}), whereas at a lower level each component density
$f_g(y|\theta_g)$ itself is described by a mixture of $H_g$ Gaussian
distributions.  On the upper level, $\Gclust$ is identified as  the number of
such clusters, whereas the number of subcomponents $H_g$ in each cluster
reflects the  quality of the semi-parametric mixture approximation.

Two different approaches are available to \lq\lq estimate\rq\rq\ the number of clusters in such a
framework.  Any such approach has to deal with
the following additional identifiability problems for this type  of mixtures:
the observed-data likelihood ascertains this\index{identifiability} model just as
one big mixture of Gaussian distributions with $\tilde{G}=  H_1+ \ldots + H_G$
components, and it does not change when we exchange subcomponents between
clusters on the lower level, even though  this leads to different cluster
distributions on the upper level of  the mixture of mixtures model.  Hence, a
mixture of mixtures model is not identifiable  in the absence of additional
information, and  this is most naturally dealt with within a Bayesian framework.

Within the Bayesian approach, it is  common to estimate the
hierarchical mixture of mixtures model directly by including such prior information;
see, in particular, \citet{mal-etal:ide} who consider a random-effects prior to
introduce prior dependence among the $H_g$ means of the subcomponent Gaussian
mixture defining $f_g(y|\theta_g)$.  A different  approach which is prevalent
in the frequentist literature employs a two-step procedure  and tries to create
meaningful clusters after having fitted a  Gaussian mixture as in
(\ref{eq:mixnormmul}) with $G=\Gmax$.  The clusters are determined by
successively merging components according to some criterion such as the
entropy of the resulting partition \citep{bau-etal:com}; see
Chapter~8 for additional approaches and further details.


\section{One-Sweep Methods for Cross-model Inference on $G$} \label{sec_bayes}

From a Bayesian perspective, inference methods that treat $G$ or  $\Gclust$  as
an unknown para\-meter to be estimated jointly with the component-specific
parameters $\thmod$ are preferable to processing $G$ as a model index and relying on
testing principles. 
Several such approaches are reviewed in this section.

\subsection{Overfitting  mixtures}  \label{sec:over}

\citet{mengersen:rousseau:2011} examine the issue of an overfitting mixture,
\index{overfitted mixtures}
 that is, the estimation of a mixture model with $G$ components when
the true distribution behind the data has fewer than $G$, say  $G_0$, components. This setting complicates even further the
non-identifiability of the mixture model, since there are ${G \choose G_0}$ ways of picking $G_0$ components out of the
$G$ (while cancelling the others); see  also Chapter~4.

\citet{mengersen:rousseau:2011}  show  that the posterior distribution
on the parameters of the overfitted mixture has a much more stable behaviour than the  likelihood function when the prior on the weights of the mixture is
sufficiently concentrated on the boundaries of the parameter space, that is, with many weights being close to  zero. In fact, the central result of
\citet{mengersen:rousseau:2011} is that, if the dimension $r$ of the component parameters is larger than twice the
hyperparameter $e_0$ of a symmetric Dirichlet prior $\Dirinv{G}{e_0}$ on the weights, then
the sum of the weights of the extra   $G-G_0$ components  asymptotically concentrates at zero.
This result has the additional appeal of validating less informative priors as asymptotically
consistent. In practice, it means that selecting a Dirichlet $\Dirinv{G}{1/2}$ and an
arbitrary prior on the component parameters should see superfluous components vanish as the sample size grows to be
large enough, even though the impact of the choice of $e_0$ can be perceived for finite sample sizes.




\subsection{Reversible jump MCMC}  \label{RJMCMC}

Reversible\index{RJMCMC}\index{reversible jump MCMC}\index{MCMC!reversible jump} jump MCMC \citep[RJMCMC;][]{green:1995} was exploited by \citet{richardson:green:1997} to select the number of components $G$ for
univariate mixtures of Gaussian distributions. As briefly discussed in  Chapter~1, this simulation method is based
on creating a Markov chain that moves over a space of variable dimensions, namely between the parameter spaces of finite
mixtures with different numbers of components, while retaining the fundamental detailed balance property that ensures
the correct stationary (posterior) distribution.

The intuition behind the RJMCMC method is to create bijections between
 pairs of parameter spaces by creating auxiliary variables that equate the dimensions of the augmented spaces and to
keep the same bijection for a move and its reverse. When designing a \mbox{RJMCMC} algorithm, those pairwise moves have to be
carefully selected in order to reach sufficiently probable regions in the new parameter space. \citet{richardson:green:1997}
discuss at length their {\em split-and-merge} moves which split (or aggregate) one (or two) components of the current
mixture, with better performance  than the basic {\em birth-and-death} moves, but performance may deteriorate as the
number of components increases.  The design of suitable proposals for higher-dimensional mixtures is quite a challenge, as demonstrated by
\citet{del-pap:mul} and \citet{zha-etal:lea} for multivariate  normal mixtures.
In an attempt to extend RJMCMC methods  to hidden Markov models, \citet{cappe:robert:ryden:2002} had to face acceptance rates as
low as 1\%.
RJMCMC is a natural extension of the traditional Metropolis--Hastings
algorithm, but calibrating it is often perceived as too great an obstacle to its implementation, and it is not competitive
with within-model simulations in the case of a small number of values of $G$ in competition.


\subsection{Allocation sampling}  \label{sec:allocation}

As discussed   in Section~\ref{sec_clust},   the main object of interest  in Bayesian clustering is the
marginal posterior of the allocations, that is, $p(\zm|\ym, G)$ (if $G$ is known) or  $p(\zm|\ym)$ (if $G$ is unknown).
Hence, Bayesian clustering has to rely on efficient methods to sample from the posterior $p(\zm|\ym, G)$ (or $p(\zm|\ym)$).

While full conditional Gibbs sampling from  the joint distribution  $p(\theta,\zm|\ym,G)$ will yield draws from the (marginal) posterior
$p(\zm|\ym, G)$,  several authors considered alternative algorithms of \lq\lq allocation sampling\rq\rq .
Early Bayesian clustering approaches without parameter estimation are  based on sampling from the marginal posterior
distribution $p(\zm|\ym,G)$, defined earlier  in (\ref{margSiv}),  for known $G$.  \citet{che-liu:pre} were among the first to show how sampling of
the allocations from $p(\zm|\ym,G)$ (for a fixed $G$)  becomes feasible  through MCMC methods, using either
single-move Gibbs sampling  or the Metropolis--Hastings algorithm; see \citet[Section~3.4]{fruhwirth:2006} and
\citet{marin:mengersen:robert:2004} for more details.

We want to stress here the following issue. Although these MCMC samplers operate in the marginal space of the allocations $\zm$,
neither  the
integrated likelihood $p(\ym |\zm, G)$,  defined earlier in (\ref{margyz}), nor the prior $p(\zm| G)$,  given  in (\ref{intro:dirich:ml}),
 can  be (properly) defined without
specifying a prior distribution $p(\theta_1,
\ldots, \theta_G, \eta |G)$  for the unknown parameters of a mixture  model with $G$ components. This problem is closely related
to the problem discussed in Section~\ref{exaICL}  of  having to choose  priors for the exact ICL  criterion.
As discussed in Chapter~4,the choice of such a prior is not obvious and may
have considerable impact on posterior
 inference. 

These early sampling algorithms focus on computational aspects and do not  explicitly account for the problem  that the number $\Gclust$ of
clusters in the sampled partitions $\zm$ might differ from $G$,  taking the identity of $G$ and $\Gclust$ more or less for  granted.
Still, as discussed above and again in Section~\ref{sec:sfm}, whether this applies or not very much depends on the
choice of the hyperparameter $\ed{0}$ in the Dirichlet prior  $\Dirinv{G}{e_0}$  on the weights.

 \citet{nob-fea:bay} address the problem of an unknown number of components $G$ in the context of
allocation sampling.  For a given $G$, they employ  the usual Dirichlet prior $\etav|G \sim \Dir(\ed{1}, \ldots,
\ed{G})$ on the weight distribution, but treat $G$ as an unknown parameter, associated with a prior $p(G)$ (e.g.\ $G-1
\sim \Poisson(1)$), 
 as justified by \citet{nob:pos}.
An MCMC sampler is
developed that draws from the joint posterior $p(\zm,G|\ym)$, by calling either Gibbs or Metropolis--Hastings moves based on the
conditional distribution of $p(\zm|\ym,G)$ for a given $G$ and by running RJMCMC type moves for switching values of $G$.
Based on these posterior draws, $\Gclusthat$ is estimated from the posterior draws of  the  number of non-empty clusters $\Gclust$.
Several post-processing strategies are discussed for solving the label switching problem that is inherent in this
sampler and for estimating $\hat{\zm}$.

\subsection{Bayesian nonparametric  methods}\label{sec:BaNo}

A quite different approach of selecting the number $\Gclust$ of clusters  exists outside the framework of finite mixture
models and relies on Bayesian   nonparametric 
approaches based on mixture models with countably infinite number of
components, as discussed in Chapter~6 in full detail.\index{Bayesian!nonparametrics}

For  Dirichlet process (DP) mixtures \citep{mue-mit:bay}, for instance,  the discrete mixing distribution in the finite
mixture (\ref{eq:mixOfMix1}) is substituted by a random distribution $H  \sim DP(\alphaDP, H_0) $, drawn from a DP prior
with precision parameter $\alphaDP$ and base measure $H_0$. As a draw $H$ from a DP is almost surely discrete, the
corresponding model has a representation as an infinite mixture,
\begin{align}
y \sim  \sum \limits_{g=1}^{\infty} \eta_g f_g(y |\theta_g), \label{eq:infMix1}
\end{align}
with \iid\ atoms $\theta_g \simiid H_0 $ drawn from the base measure
$H_0$ and  weights $\eta_g$ obeying the stick-breaking representation
\begin{align}
 \eta_g=\stick_g \prod_{j=1}^{g-1} (1- \stick_j ),  \quad g=1, 2,\ldots, \label{stickr}
\end{align}
with $\stick_g \sim \Betadis(1,\alphaDP) $ \citep{set:con}.

As DP priors induce ties among the observations,
such an  approach  automatically induces a random partition (or\index{random!partition!models}
clustering) $\parti$ of  the data
with a corresponding random cardinality $\Gclust$ (see Chapter~6).
Since there are infinitely many
components in  (\ref{eq:infMix1}) (i.e.\ $G=\infty$), there is no risk of confusing $G$ and $\Gclust$  as  for  finite mixtures.
 For a DP prior with precision parameter $\alphaDP$, the prior distribution over the partitions $\parti$ is given by 
\begin{align} \label{priparti}
\displaystyle p(\parti|\alphaDP,\Gclust)=  \alphaDP^{\Gclust}  \frac{\Gamfun{\alphaDP} }{ \Gamfun{n+ \alphaDP}} \prod_{g:n_g>0}
 \Gamfun{n_g},
 \end{align}
 where  $n_g$ and  $\Gclust$ are defined as in (\ref{intro:dirich:ml}).
 Another defining property of the DP prior is the structure of the prior predictive distribution
$p(z_i|\zm_{-i})$, where $\zm_{-i}$ denotes all indicators excluding $z_i$.  Let $\Kni$ be the number of non-empty
clusters implied by $\zm_{-i}$ and let $\Nki{g}$, $g=1, \ldots, \Kni$, be the corresponding cluster sizes.
Then the  probability
 that $z_i$ is assigned to an existing cluster $g$ is given by
\begin{eqnarray}
\displaystyle \Proba{z_i=g|\zm_{-i}, \Nki{g} >0} 
 = \frac{\Nki{g} }{n-1 +  \alphaDP}, \label{seqDP}
\end{eqnarray}
whereas  the prior probability that  $z_i$ creates a new cluster (indexed by $\Kni +1 $) is equal to
\begin{eqnarray}
\displaystyle \Proba{z_i = \Kni +1 |\zm_{-i}} =  \frac{\alphaDP}{n-1 + \alphaDP}.  \label{seq_emptyDP}
\end{eqnarray}
Given this strong focus on  BNP mixtures as random partition models, it is not surprising  that the main interest in posterior inference is
again in the draws from the posterior $p(\zm|\ym)$  of the allocations which  are exploited in various ways  to  choose an
appropriate  partition $\hat{\zm}$ of the data and to estimate the  number of clusters $\Gclust$.

\citet{lau-gre:bay} compare  BNP methods to estimate the number of
clusters   with the outcome associated with finite mixtures.  They
also show in detail how to derive a single (optimal) point estimate $\hat{\zm}$  from the posterior  $p(\zm|\ym)$, with
the number of distinct  clusters $\Gclusthat$  in $\hat{\zm}$ being an estimator of $\Gclust$ in this framework.
   To derive a partition
of the data, \citet{mol-etal:bay} cluster  the data using
 the pairwise association matrix as a distance measure  which is obtained by aggregating
over all partitions obtained during MCMC sampling, using
partitioning around medoids. 
The optimal number  of clusters is determined by maximizing an associated clustering
score; see also \citet{Mix:Liverani2013}.

\begin{figure}[t!]
\begin{center}
\scalebox{0.35}{\includegraphics{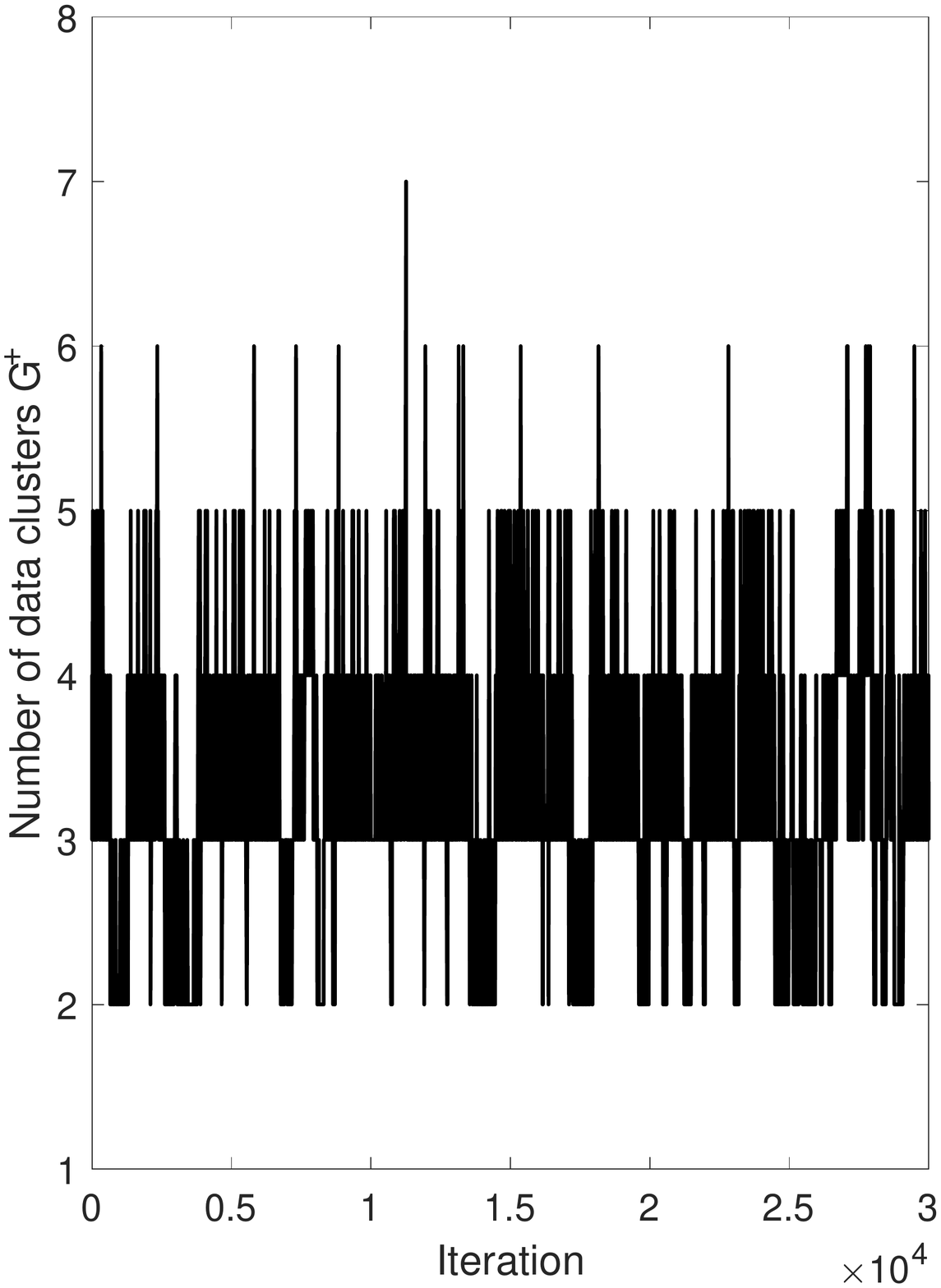}}
\scalebox{0.35}{\includegraphics{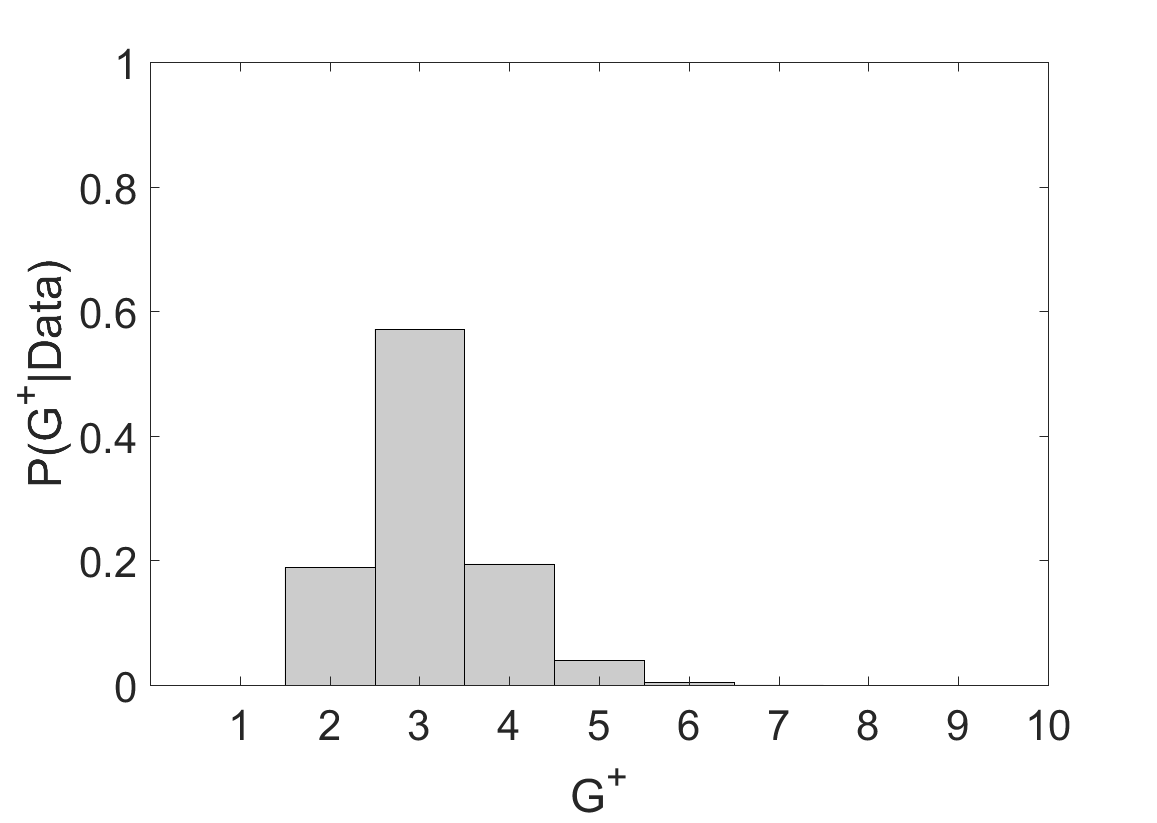}}
\end{center}
\caption{Sparse finite mixture modelling of the enzyme data:
(left) 30,000 posterior draws of the number of data clusters $\Gclust$;
(right) posterior distribution $p(\Gclust|\ym)$.}\label{cfr:fig4}
\end{figure}

A well-known limitation of DP priors is that \textit{a~priori} the cluster sizes are expected to be geometrically ordered, with one big cluster,
geometrically smaller clusters, and many singleton clusters \citep{mue-mit:bay}.  This initiated the investigation of alternative BNP mixtures and their
usefulness for clustering.  A popular BNP two-parameter mixture is obtained from the Pitman--Yor process (PYP) prior $PY{(\betaPY,\alphaPY)}$
with $\betaPY \in [0,1)$, $\alphaDP > -\betaPY $  \citep{pit-yor:two}, with a stick-breaking representation\index{Pitman--Yor!process}
as in (\ref{stickr}) with
$\stick_g \sim \Betadis(1-\betaPY,\alphaDP+k\betaPY)$. The DP prior occurs as a special case when $\betaPY=0$.
PYP mixtures are known to be more
useful than the DP mixture for data with many significant, but small clusters. 

For a DP as well as a PYP mixture,  the prior expected number of data
clusters $\Kn$ increases as the number $n$ of observations increases,
where  for the DP process $\Kn \sim \alphaDP \log (n)$  \citep{kor-hol:con}
and $\Kn \sim n ^ \betaPY$ obeys a power law for PYP mixtures.
As will be discussed in the next subsection, 
 finite mixtures are quite different in this respect.

\subsection{Sparse finite mixtures for model-based clustering}  \label{sec:sfm}

Inspired by the  important insights of \citet{mengersen:rousseau:2011}, \citet{MWFSG16}  introduced the concept  of
sparse finite mixture models for model-based clustering  as an alternative to infinite mixtures, 
following ideas presented earlier  in \citet{SFR11}. A similar approach is pursued by \citet{van-etal:ove}.

 While remaining within the framework of finite mixtures, sparse\index{mixtures!sparse}
finite mixture models provide a semi-parametric Bayesian approach in so far as  the number $\Gclust$ of non-empty mixture components
used to generate the data is not assumed to be known in advance, but random, as already discussed in Section~\ref{sec_clust}.
The basic idea of sparse finite mixture modelling is to deliberately specify an
\textit{overfitting} finite mixture model with too many components $G$.
Sparse finite mixtures  stay within the common finite mixture framework by
assuming a symmetric Dirichlet prior $\etav \sim \Dirinv{G}{\ed{0}}$  on the weight distribution;
however, the hyperparameter $e_0$ of this prior is selected such that superfluous
components are emptied automatically during MCMC sampling and
sparse solutions with regard to the number  $\Gclust$ of clusters
 are induced  through the prior on the weight distribution.
 This proposal leads to a simple Bayesian
framework where a straightforward MCMC 
sampling procedure is applied to jointly estimate the
unknown number of non-empty data clusters  $\Gclust$ together with the remaining parameters.


As discussed in Section~\ref{sec_clust}, for such a mixture model,  the number $G$ of components does  not reflect the
number of data clusters, as many components will remain unused.  Following  \citet{nob:pos}, \citet{MWFSG16} derive
the posterior distribution  $\Proba{\Gclust=g|\ym},\ g=1,\ldots, G,$ of the number $\Gclust$ of data clusters from
the MCMC output of the allocations $\zm$. Therefore, for each iteration $m$ of  MCMC sampling,  all components $g$ to
which some observations have been assigned  are identified from  $\zm \im{m}$ and the corresponding  number of  non-empty
components  is considered:
$$
\Gclust  \im{m}=G-\sum \limits_{g=1}^G \indic{n_g \im{m} =0}, \label{eq:K0}
$$
where, for $g=1,\ldots,G$,  $n_g \im{m}= \sum_{i=1}^n \indic{z_i\im{m} =g} $ is the number of observations
allocated to component $g$, and $\indic{\cdot}$ denotes the indicator function.  The posterior  distribution
$\Proba{\Gclust=g|\ym}$, $g=1,\ldots, G,$ is then estimated by the corresponding relative frequency.

The number of  clusters $\Gclust$ can be  derived  as a point estimator from this distribution, for example, the posterior mode
estimator $\Gclusttilde$  that maximizes the (estimated) posterior distribution $\Proba{\Gclust=g|\ym}$. This happens to
be the most frequent number of clusters  visited during MCMC sampling. The posterior mode estimator appears
to be sensible in the present context when adding very small clusters hardly changes the marginal likelihood.
This  makes the posterior distribution $\Proba{\Gclust=g|\ym}$ extremely right-skewed,  and other point
estimators such as the posterior mean are extremely sensitive to prior choices, as  noted by \citet{nob:pos}. However, under a framework
where sparse finite mixtures are employed  for density estimation, very small components might be important  and other
estimators of  $\Gclust$ might  be better justified.

 An alternative way to summarize clustering based on sparse finite mixtures is by exploring the posterior draws of the partitions $\zm$
 and determining some optimal  partition, such as the partition
  $\hat{\zm}$ minimizing Binder's loss function. This can be done without the need to resolve label switching or to stratify
 the draws with respect to $\Gclust$.  The cardinality $\Gclusthat$  of  such an optimal partition  $\hat{\zm}$ is yet another estimator of the number of
clusters.  The posterior mode estimator $\Gclusttilde$ and $\Gclusthat$ do  not necessarily coincide, and differences in these estimators  reflect   uncertainty
 in the posterior distribution
over the partition space.   As discussed by \citet{fru-etal:con},
the approach of \cite{wad-gha:bay}  to quantifying such uncertainty can be applied immediately to sparse finite mixture models.

\begin{figure}[t!]
\begin{center}
\scalebox{0.5}{\includegraphics{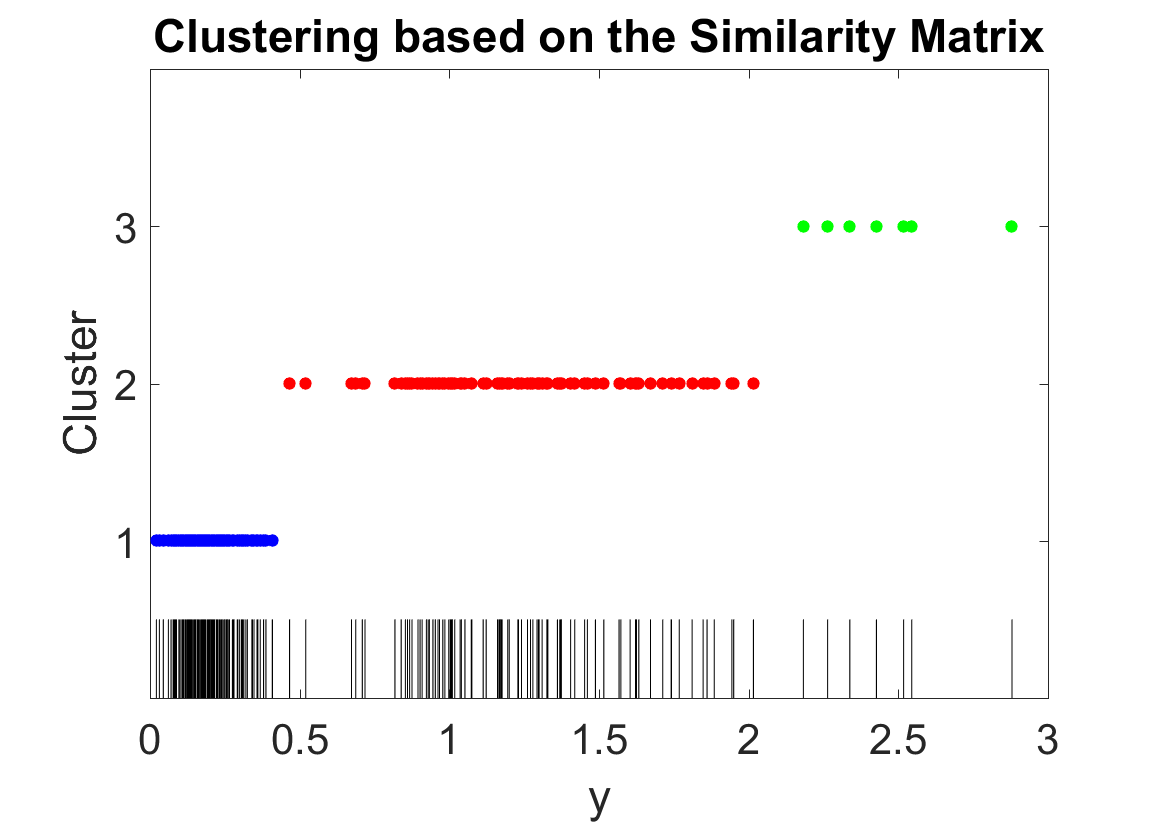}}
\end{center}
\caption{Sparse finite mixture modelling of the enzyme data,  displayed as a rug plot. Partition $\hat{\zm}$ optimizing Binder's loss function.
The number of clusters in this partition is equal to three.}\label{cfr:fig5}
\end{figure}

  %
  %

 The appropriate choice of the hyperparameter $e_0$ is important for the application of
the sparse finite mixture approach in a clustering context. While in a density estimation framework
the asymptotic criterion  of \citet{mengersen:rousseau:2011}  suggests the choice $ e_0 < r/2$,
with $r$ being the dimension of $\theta_g$,
this rule is not necessarily a sensible  choice for  selecting the  number of clusters  $\Gclust$  in a data set of finite size $n$,
 as   demonstrated for a broad range of mixture models in \citet{MWFSG16} and \citet{fru-mal:fro}.
  Indeed,  
  these papers\index{number of components!versus number of clusters}
 show that   values of $e_0  \ll r/2$ much smaller  than  the  asymptotic criterion of  \citet{mengersen:rousseau:2011}  are needed
to identify the right number of clusters, and recommend choosing either very small fixed values such as  $e_0  = 0.001$ or applying a hyperprior with
$ e_0 \sim \Gammad(a_e ,b_e)$ such that $\Ew(e_0)= a_e/b_e$ is very small (e.g.\ $ e_0 \sim \Gammad(1 ,200)$).

Under the provision that $G_+$ underestimates $G$, this approach constitutes a simple and  generic strategy for model selection
 without making use of model selection criteria, \mbox{RJMCMC},  or marginal
likelihoods. Applications include Gaussian mixtures as well as mixtures of Gaussian mixtures
\citep{mal-etal:ide} and sparse mixtures for discrete-valued data  \citep{fru-mal:fro}.
 By way of further illustration,  the enzyme data (shown earlier in Figure~\ref{cfr:fig1})  are reanalysed using sparse finite mixtures,
 taking  the prior of \citet{richardson:green:1997} as base measure.
The maximum number of data clusters is chosen as $G=10$ and the hierarchical sparse Dirichlet prior
$\etav \sim \Dirinv{G}{\ed{0}}$,  $ e_0 \sim \Gammad(1 ,200)$ is applied.

Figure~\ref{cfr:fig4}  shows 30,000 posterior draws of the number of data clusters $\Gclust$ as well as
the corresponding posterior distribution $p(\Gclust|\ym)$.  The posterior mode estimator yields  three clusters with $\Proba{\Gclust=3|\ym}=0.57$.
 Also two  clusters are supported with   $\Proba{\Gclust=2|\ym}=0.19$, which
 is not unexpected in the light of Figure~\ref{cfr:fig1}, showing two (albeit non-Gaussian) data clusters.
Due to this misspecification of  the component densities  the four-cluster solution is equally supported with $\Proba{\Gclust=4|\ym}=0.19$.
 Finally, Figure~\ref{cfr:fig5} shows the partition $\hat{\zm}$ optimizing Binder's loss function together with a rug plot of the data. The number of clusters in this
partition is equal to three, supporting the choice based on the posterior mode. The resulting clustering nicely captures the three distinct groups of data points.

\paragraph*{Relation to BNP methods}

The concept of sparse finite mixtures  is related in  various ways to DP 
mixtures, discussed in Section~\ref{sec:BaNo}.
 If  the weight distribution  follows the Dirichlet prior $\etav \sim \Dir_G(\alphaDP/G)$
  and the base measure  $H_0$ serves as  prior for  the component parameters
 (i.e. $\theta_g \sim H_0 $),  then, as shown by \citet{gre-ric:mod},   the  finite mixture in  (\ref{eq:mixOfMix1}) converges to a DP
mixture with mixing distribution $H \sim DP(\alphaDP,H_0)$ as $G$ increases.
This relationship has mainly been exploited to obtain a finite mixture approximation to the  DP mixture.
 In this sense, the sparse finite Gaussian mixture introduced in \citet{MWFSG16}  could be seen as an approximation
to a DP mixture. 
Nevertheless, as argued by \citet{mal-etal:ide}, it makes sense to stay within the framework of finite mixtures
and to consider  $G$ as a second parameter which is held fixed at  a finite value, as this provides a two-parameter
alternative to  DP  mixtures with related properties.

Representations  similar   to  BNP mixtures  exist also  for finite mixture models under the symmetric
prior $\etav \sim \Dirinv{G}{\ed{0}}$,
but  are not commonly known, although they  shed further light on the relation between the two model classes.
First of all,  a stick-breaking representation  of the weights  $\eta_1, \eta_2,  \ldots, \eta_G $ as in (\ref{stickr})
  in terms of a sequence of independently (albeit not identically) distributed random variables  exists 
 also for finite mixtures, with
$\stick_g \sim \Betadis (\ed{0},(G-g)\ed{0})$, $g=1,\ldots, G-1$,  $\stick_G=1$; see, for example, \citet{SFR11}.

Second, as already discussed in Section~\ref{sec:exp_par}, finite mixture models can be regarded as random partition models  and
 the prior distribution over all random partitions $\parti$ of $n$ observations
can be  derived from the joint (marginal) prior  $ p(\zm|G) $   given
in (\ref{intro:dirich:ml}) (see, for example, \citet{mal-etal:ide}):
\begin{eqnarray*}
\displaystyle p(\parti| \ed{0}, \Gclust) = \frac{G!}{(G-\Gclust) !}
 \frac{\Gamfun{G \ed{0}} }{\Gamfun{n+ G \ed{0}}
 \Gamfun{\ed{0}}^{\Kn}} \prod_{g: n_g>0 } \Gamfun{n_g+\ed{0}}.  \label{ijhuh:ml}
\end{eqnarray*}
This prior 
takes the form of a product partition model as for DP mixtures (see (\ref{priparti}))  and  is invariant to permuting the cluster labels.

Finally, as for BNP mixtures,  it is possible to derive the prior predictive distribution
$p(z_i|\zm_{-i})$, where $\zm_{-i}$ denotes all indicators excluding $z_i$.  Let $\Kni$ be the number of non-empty
clusters implied by $\zm_{-i}$, and let $\Nki{g}$, $g=1, \ldots, \Kni$, be the corresponding cluster sizes.
Then the  probability
 that $z_i$ is assigned to an existing cluster $g$ is given by 
$$
\displaystyle \Proba{z_i=g|\zm_{-i}, \Nki{g} >0} 
 = \frac{\Nki{g} + \ed{0}}{n-1 + \ed{0} G}, \label{seq}
$$
which is closely related to (\ref{seqDP}), in particular if $e_0=\alpha/G$ and $G$ increases.
However,  the prior probability that  $z_i$ creates a new cluster with $z_i \in I =\{g :  \Nki{g}=0 \} $  is equal to
\begin{eqnarray}
\displaystyle \Proba{z_i \in I |\zm_{-i}} = 
\frac{\ed{0} (G-\Kni)}{n-1 + \ed{0} G},  \label{seq_empty}
\end{eqnarray}
 and is  quite different from (\ref{seq_emptyDP}).
In particular, for $\ed{0}$  independent of $G$, this probability not only depends on $\ed{0}$, but also increases with $G$. Hence a sparse finite
mixture model  can be regarded as a two-parameter model, where both
$\ed{0}$ and $G$ influence the prior expected number of data clusters $\Kn$, which is determined for a DP mixture solely
by $\alphaDP$. Furthermore, the prior probability (\ref{seq_empty}) of creating new clusters decreases as
the number $\Kni$ of non-empty clusters increases, as opposed to  DP mixtures where this probability is
constant and to PYP mixtures where this probability increases.
Hence,  sparse finite mixtures  are useful for clustering  data that arise from a moderate number of clusters that does not increase as the number of
data points $n$ increases.

\section{Concluding Remarks}

The issue of selecting the number of mixture components has always been
contentious, both in frequentist and Bayesian terms, and this chapter has
reflected on this issue by presenting a wide variety of solutions and analyses.
The main reason for the difficulty in estimating the order $G$ of a mixture
model is that it is a poorly defined quantity, even when setting aside
identifiability and label switching aspects. Indeed, when considering a single
sample of size $n$ truly generated from a finite mixture model, there is always
a positive probability that the observations in that sample are generated from
a subset of the components of the mixture of size $G_+$  rather than from all components
$G$. As shown by the asymptotic results in Chapter~4, the issue goes away as the sample size $n$
goes to infinity (provided $G$ remains fixed), but this does not bring a
resolution to the quandary of whether or not $G$ is estimable. In our opinion,
inference should primarily bear on the number of data clusters $G_+$, since the
conditional posterior distribution of $G$ given $G_+$ mostly depends on the
prior modelling and very little on the data. Without concluding like Larry
Wasserman (on his now defunct {\em Normal Deviate} blog) that ``mixtures, like
tequila,\index{tequila} are inherently evil and should be avoided at all
costs'', we must acknowledge that the multifaceted uses of mixture models
imply that the estimation of a quantity such as the number of mixture components
should be impacted by the purpose of modelling via finite mixtures, as for
instance through the prior distributions in a Bayesian setting.

\hyphenation{Post-Script Sprin-ger}\hyphenation{Post-Script Sprin-ger}


\begin{thebibliography}{98}
\expandafter\ifx\csname natexlab\endcsname\relax\def\natexlab#1{#1}\fi

\bibitem[{Akaike(1974)}]{Akaike}
\textsc{Akaike, H.} (1974).
\newblock A new look at the statistical model identification.
\newblock \textit{IEEE Transactions on Automatic Control} \textbf{19},
  716--723.

\bibitem[{Banfield \& Raftery(1993)}]{ban-raf:mod}
\textsc{Banfield, J.~D.} \& \textsc{Raftery, A.~E.} (1993).
\newblock Model-based {G}aussian and non-{G}aussian clustering.
\newblock \textit{Biometrics} \textbf{49}, 803--821.

\bibitem[{Baudry(2015)}]{Baudry15}
\textsc{Baudry, J.-P.} (2015).
\newblock Estimation and model selection for model-based clustering with the
  conditional classification likelihood.
\newblock \textit{Electronic Journal of Statistics} \textbf{9}, 1041--1077.

\bibitem[{Baudry et~al.(2012)Baudry, Maugis \& Michel}]{BMM12}
\textsc{Baudry, J.-P.}, \textsc{Maugis, C.} \& \textsc{Michel, B.} (2012).
\newblock Slope heuristics: Overview and implementation.
\newblock \textit{Statistics and Computing} \textbf{22}, 455--470.

\bibitem[{Baudry et~al.(2010)Baudry, Raftery, Celeux, Lo \&
  Gottardo}]{bau-etal:com}
\textsc{Baudry, J.-P.}, \textsc{Raftery, A.~E.}, \textsc{Celeux, G.},
  \textsc{Lo, K.} \& \textsc{Gottardo, R.} (2010).
\newblock Combining mixture components for clustering.
\newblock \textit{Journal of Computational and Graphical Statistics}
  \textbf{19}, 332--353.

\bibitem[{Berger(1985)}]{berger:1985}
\textsc{Berger, J.} (1985).
\newblock \textit{Statistical Decision Theory and {B}ayesian Analysis}.
\newblock New York: Springer-Verlag, 2nd ed.

\bibitem[{Berger \& Jefferys(1992)}]{berger:jefferys:1992}
\textsc{Berger, J.} \& \textsc{Jefferys, W.} (1992).
\newblock Sharpening {O}ckham's razor on a {B}ayesian strop.
\newblock \textit{American Statistician} \textbf{80}, 64--72.

\bibitem[{Berkhof et~al.(2003)Berkhof, {van Mechelen} \&
  Gelman}]{berkhof:mechelen:gelman:2003}
\textsc{Berkhof, J.}, \textsc{{van Mechelen}, I.} \& \textsc{Gelman, A.}
  (2003).
\newblock A {B}ayesian approach to the selection and testing of mixture models.
\newblock \textit{Statistica Sinica} \textbf{13}, 423--442.

\bibitem[{Bertoletti et~al.(2015)Bertoletti, Friel \& Rastelli}]{BFR15}
\textsc{Bertoletti, M.}, \textsc{Friel, N.} \& \textsc{Rastelli, R.} (2015).
\newblock Choosing the number of components in a finite mixture using an exact
  integrated completed likelihood criterion.
\newblock \textit{Metron} \textbf{73}, 177--199.

\bibitem[{Besag(1989)}]{bes:can}
\textsc{Besag, J.} (1989).
\newblock A candidate's formula: {A} curious result in {B}ayesian prediction.
\newblock \textit{Biometrika} \textbf{76}, 183.

\bibitem[{Biernacki et~al.(2000)Biernacki, Celeux \& Govaert}]{BCG00}
\textsc{Biernacki, C.}, \textsc{Celeux, G.} \& \textsc{Govaert, G.} (2000).
\newblock Assessing a mixture model for clustering with the integrated
  completed likelihood.
\newblock \textit{{IEEE} Transactions on Pattern Analysis and Machine
  Intelligence} \textbf{22}, 719--725.

\bibitem[{Biernacki et~al.(2010)Biernacki, Celeux \& Govaert}]{BCG10}
\textsc{Biernacki, C.}, \textsc{Celeux, G.} \& \textsc{Govaert, G.} (2010).
\newblock Exact and {M}onte {C}arlo calculations of integrated likelihoods for
  the latent class model.
\newblock \textit{Journal of Statistical Planning Inference} \textbf{140},
  2991--3002.

\bibitem[{Binder(1978)}]{bin:bay}
\textsc{Binder, D.~A.} (1978).
\newblock Bayesian cluster analysis.
\newblock \textit{Biometrika} \textbf{65}, 31--38.

\bibitem[{Birg\'e \& Massart(2001)}]{BirgeMassart01}
\textsc{Birg\'e, L.} \& \textsc{Massart, P.} (2001).
\newblock Gaussian model selection.
\newblock \textit{Journal of the European Mathematical Society} \textbf{3},
  203--268.

\bibitem[{Birg\'e \& Massart(2007)}]{BirgeMassart07}
\textsc{Birg\'e, L.} \& \textsc{Massart, P.} (2007).
\newblock Gausian model selection.
\newblock \textit{Probability Theory and Related Fields} \textbf{138}, 33--73.

\bibitem[{Bouveyron et~al.(2015)Bouveyron, C\^ome \& Jacques}]{BCJ15}
\textsc{Bouveyron, C.}, \textsc{C\^ome, E.} \& \textsc{Jacques, J.} (2015).
\newblock The discriminative functional mixture model for a comparative
  analysis of bike sharing systems.
\newblock \textit{Annals of Applied Statistics} \textbf{9}, 1726--1760.

\bibitem[{Box(1976)}]{Box}
\textsc{Box, G. E.~P.} (1976).
\newblock An application of the {L}aplace method to finite mixture
  distributions.
\newblock \textit{Journal of the American Statistical Association} \textbf{71},
  791--799.

\bibitem[{Bozdogan(1987)}]{Bozdogan87}
\textsc{Bozdogan, H.} (1987).
\newblock Model selection and {A}kaike's information criterion ({AIC}): The
  general theory and its analytical extensions.
\newblock \textit{Psychometrika} \textbf{52}, 345--370.

\bibitem[{Capp\'e et~al.(2002)Capp\'e, Robert \&
  Ryd\'en}]{cappe:robert:ryden:2002}
\textsc{Capp\'e, O.}, \textsc{Robert, C.~P.} \& \textsc{Ryd\'en, T.} (2002).
\newblock Reversible jump {MCMC} converging to birth-and-death {MCMC} and more
  general continuous time samplers.
\newblock \textit{Journal of the Royal Statistical Society, Series B}
  \textbf{65}, 679--700.

\bibitem[{Casella et~al.(2004)Casella, Robert \& Wells}]{cas-etal:mix}
\textsc{Casella, G.}, \textsc{Robert, C.~P.} \& \textsc{Wells, M.~T.} (2004).
\newblock Mixture models, latent variables and partitioned importance sampling.
\newblock \textit{Statistical Methodology} \textbf{1}, 1--18.

\bibitem[{Celeux et~al.(2001)Celeux, Chr\'etien, Forbes \& Mkhadri}]{CCFM}
\textsc{Celeux, G.}, \textsc{Chr\'etien, S.}, \textsc{Forbes, F.} \&
  \textsc{Mkhadri, A.} (2001).
\newblock A component-wise {EM} algorithm for mixtures.
\newblock \textit{Journal of Computational and Graphical Statistics}
  \textbf{10}, 697--712.

\bibitem[{Celeux et~al.(2006)Celeux, Forbes, Robert \&
  Titterington}]{cel-etal:dev}
\textsc{Celeux, G.}, \textsc{Forbes, F.}, \textsc{Robert, C.~P.} \&
  \textsc{Titterington, D.~M.} (2006).
\newblock Deviance information criteria for missing data models.
\newblock \textit{Bayesian Analysis} \textbf{1}, 651--674.

\bibitem[{Celeux \& Soromenho(1996)}]{Celeux96}
\textsc{Celeux, G.} \& \textsc{Soromenho, G.} (1996).
\newblock An entropy criterion for assessing the number of clusters in a
  mixture model.
\newblock \textit{Journal of Classification} \textbf{13}, 195--212.

\bibitem[{Chen \& Liu(1996)}]{che-liu:pre}
\textsc{Chen, R.} \& \textsc{Liu, J.~S.} (1996).
\newblock Predictive updating methods with application to {B}ayesian
  classification.
\newblock \textit{Journal of the Royal Statistical Society, Series B}
  \textbf{58}, 397--415.

\bibitem[{Chib(1995)}]{chib:1995}
\textsc{Chib, S.} (1995).
\newblock Marginal likelihood from the {G}ibbs output.
\newblock \textit{Journal of the American Statistical Association} \textbf{90},
  1313--1321.

\bibitem[{Chopin(2002)}]{chopin:2002}
\textsc{Chopin, N.} (2002).
\newblock A sequential particle filter method for static models.
\newblock \textit{Biometrika} \textbf{89}, 539--552.

\bibitem[{Chopin \& Robert(2010)}]{chopin:robert:2010}
\textsc{Chopin, N.} \& \textsc{Robert, C.~P.} (2010).
\newblock Properties of nested sampling.
\newblock \textit{Biometrika} \textbf{97}, 741--755.

\bibitem[{C\^ome \& Latouche(2015)}]{Come15}
\textsc{C\^ome, E.} \& \textsc{Latouche, P.} (2015).
\newblock Model selection and clustering in stochastic block models based on
  the exact integrated complete data likelihood.
\newblock \textit{Statistical Modelling} \textbf{6}, 564--589.

\bibitem[{Dacunha-Castelle \& Gassiat(1997)}]{dacunha:gassiat:1997b}
\textsc{Dacunha-Castelle, D.} \& \textsc{Gassiat, E.} (1997).
\newblock The estimation of the order of a mixture model.
\newblock \textit{Bernoulli} \textbf{3}, 279--299.

\bibitem[{Dacunha-Castelle \& Gassiat(1999)}]{dacunha:gassiat:1999}
\textsc{Dacunha-Castelle, D.} \& \textsc{Gassiat, E.} (1999).
\newblock Testing the order of a model using locally conic parametrization:
  Population mixtures and stationary {ARMA} processes.
\newblock \textit{Annals of Statistics} \textbf{27}, 1178--1209.

\bibitem[{De~Iorio \& Robert(2002)}]{deiorio:robert:2002}
\textsc{De~Iorio, M.} \& \textsc{Robert, C.~P.} (2002).
\newblock Discussion of ``{B}ayesian measures of model complexity and fit'', by
  {S}piegelhalter et al.
\newblock \textit{Journal of the Royal Statistical Society, Series B}
  \textbf{64}, 629--630.

\bibitem[{Dellaportas \& Papageorgiou(2006)}]{del-pap:mul}
\textsc{Dellaportas, P.} \& \textsc{Papageorgiou, I.} (2006).
\newblock Multivariate mixtures of normals with unknown number of components.
\newblock \textit{Statistics and Computing} \textbf{16}, 57--68.

\bibitem[{DiCiccio et~al.(1997)DiCiccio, Kass, Raftery \&
  Wasserman}]{dic-etal:com}
\textsc{DiCiccio, T.~J.}, \textsc{Kass, R.~E.}, \textsc{Raftery, A.} \&
  \textsc{Wasserman, L.} (1997).
\newblock Computing {B}ayes factors by combining simulation and asymptotic
  approximations.
\newblock \textit{Journal of the American Statistical Association} \textbf{92},
  903--915.

\bibitem[{Drton \& Plummer(2017)}]{DrtonPlummer2017}
\textsc{Drton, M.} \& \textsc{Plummer, M.} (2017).
\newblock A {B}ayesian information criterion for singular models.
\newblock \textit{Journal of the Royal Statistical Society, Series B}
  \textbf{79}, 1--18.

\bibitem[{Figueiredo \& Jain(2002)}]{Figueiredo02}
\textsc{Figueiredo, M.} \& \textsc{Jain, A.~K.} (2002).
\newblock Unsupervised learning of finite mixture models.
\newblock \textit{IEEE Transactions on Pattern Analysis and Machine
  Intelligence} \textbf{24}, 381--396.

\bibitem[{Fraley \& Raftery(2002)}]{Fraley02}
\textsc{Fraley, C.} \& \textsc{Raftery, A.~E.} (2002).
\newblock Model-based clustering, discriminant analysis, and density
  estimation.
\newblock \textit{Journal of the American Statistical Association} \textbf{97},
  611--631.

\bibitem[{Fr{\"u}hwirth-Schnatter(1995)}]{fru:bay}
\textsc{Fr{\"u}hwirth-Schnatter, S.} (1995).
\newblock Bayesian model discrimination and {B}ayes factors for linear
  {G}aussian state space models.
\newblock \textit{Journal of the Royal Statistical Society, Series B}
  \textbf{57}, 237--246.

\bibitem[{Fr{\"u}hwirth-Schnatter(2001)}]{fru:mcm}
\textsc{Fr{\"u}hwirth-Schnatter, S.} (2001).
\newblock Markov chain {M}onte {C}arlo estimation of classical and dynamic
  switching and mixture models.
\newblock \textit{Journal of the American Statistical Association} \textbf{96},
  194--209.

\bibitem[{Fr{\"u}hwirth-Schnatter(2004)}]{fruhwirth:2004}
\textsc{Fr{\"u}hwirth-Schnatter, S.} (2004).
\newblock Estimating marginal likelihoods for mixture and {M}arkov switching
  models using bridge sampling techniques.
\newblock \textit{Econometrics Journal} \textbf{7}, 143--167.

\bibitem[{Fr{\"u}hwirth-Schnatter(2006)}]{fruhwirth:2006}
\textsc{Fr{\"u}hwirth-Schnatter, S.} (2006).
\newblock \textit{Finite Mixture and Markov Switching Models}.
\newblock New York: Springer-Verlag.

\bibitem[{Fr{\"u}hwirth-Schnatter(2011)}]{SFR11}
\textsc{Fr{\"u}hwirth-Schnatter, S.} (2011).
\newblock Dealing with label switching under model uncertainty.
\newblock In \textit{Mixture {E}stimation and {A}pplications}, K.~Mengersen,
  C.~P. Robert \& D.~Titterington, eds., chap.~10. Chichester: Wiley, pp.
  193--218.

\bibitem[{Fr{\"u}hwirth-Schnatter(2018)}]{fru:bayesf}
\textsc{Fr{\"u}hwirth-Schnatter, S.} (2018).
\newblock \textit{Applied Bayesian Mixture Modelling. {I}mplementations in
  MATLAB using the package bayesf Version 4.0}.
\newblock
  \textsf{https://www.wu.ac.at/statmath/faculty-staff/faculty/sfruehwirthschnatter/}.

\bibitem[{Fr{\"u}hwirth-Schnatter et~al.(2018)Fr{\"u}hwirth-Schnatter, Gr{\"u}n
  \& Malsiner-Walli}]{fru-etal:con}
\textsc{Fr{\"u}hwirth-Schnatter, S.}, \textsc{Gr{\"u}n, B.} \&
  \textsc{Malsiner-Walli, G.} (2018).
\newblock Contributed comment on article by {W}ade and {G}haramani.
\newblock \textit{Bayesian Analysis} \textbf{13}, 601--603.

\bibitem[{Fr{\"u}hwirth-Schnatter \& Malsiner-Walli(2018)}]{fru-mal:fro}
\textsc{Fr{\"u}hwirth-Schnatter, S.} \& \textsc{Malsiner-Walli, G.} (2018).
\newblock From here to infinity -- sparse finite versus {D}irichlet process
  mixtures in model-based clustering.
\newblock Preprint, arXiv:1706.07194.

\bibitem[{Fr{\"u}hwirth-Schnatter \& Pyne(2010)}]{fru-pyn:bay}
\textsc{Fr{\"u}hwirth-Schnatter, S.} \& \textsc{Pyne, S.} (2010).
\newblock {Bayesian inference for finite mixtures of univariate and
  multivariate skew normal and skew-$t$ distributions}.
\newblock \textit{Biostatistics} \textbf{11}, 317--336.

\bibitem[{Gassiat(2002)}]{gassiat:2002}
\textsc{Gassiat, E.} (2002).
\newblock Likelihood ratio inequalities with applications to various mixtures.
\newblock \textit{Annales de l'Institut Henri Poincar\'e -- Probabilit\'es et
  Statistiques} \textbf{38}, 887--906.

\bibitem[{Gassiat \& van Handel(2013)}]{GassiatHandel}
\textsc{Gassiat, E.} \& \textsc{van Handel, R.} (2013).
\newblock Consistent order estimation and minimal penalties.
\newblock \textit{IEEE Transactions on Information Theory} \textbf{59},
  1115--1128.

\bibitem[{Gelfand \& Dey(1994)}]{gelfand:dey:1994}
\textsc{Gelfand, A.} \& \textsc{Dey, D.} (1994).
\newblock {B}ayesian model choice: Asymptotics and exact calculations.
\newblock \textit{Journal of the Royal Statistical Society, Series B}
  \textbf{56}, 501--514.

\bibitem[{Gelfand \& Smith(1990)}]{gelfand:smith:1990}
\textsc{Gelfand, A.} \& \textsc{Smith, A.} (1990).
\newblock Sampling based approaches to calculating marginal densities.
\newblock \textit{Journal of the American Statistical Association} \textbf{85},
  398--409.

\bibitem[{Gelman \& Meng(1998)}]{gelman:meng:1998}
\textsc{Gelman, A.} \& \textsc{Meng, X.} (1998).
\newblock Simulating normalizing constants: From importance sampling to bridge
  sampling to path sampling.
\newblock \textit{Statistical Science} \textbf{13}, 163--185.

\bibitem[{Green(1995)}]{green:1995}
\textsc{Green, P.~J.} (1995).
\newblock Reversible jump {MCMC} computation and {B}ayesian model
  determination.
\newblock \textit{Biometrika} \textbf{82}, 711--732.

\bibitem[{Green \& Richardson(2001)}]{gre-ric:mod}
\textsc{Green, P.~J.} \& \textsc{Richardson, S.} (2001).
\newblock Modelling heterogeneity with and without the {D}irichlet process.
\newblock \textit{Scandinavian Journal of Statistics} \textbf{28}, 355--375.

\bibitem[{Hansen(1992)}]{Hansen92}
\textsc{Hansen, B.} (1992).
\newblock The likelihood ratio test under non-standard conditions: {T}esting
  the {M}arkov switching model of {GNP}.
\newblock \textit{Journal of Applied Econometrics} \textbf{7}, S61--S82.

\bibitem[{Keribin(2002)}]{Keribin}
\textsc{Keribin, C.} (2002).
\newblock Consistent estimation of the order of mixture models.
\newblock \textit{Sankhy\={a}, Series A} \textbf{62}, 49--66.

\bibitem[{Keribin et~al.(2015)Keribin, Brault, Celeux \& Govaert}]{KBCG15}
\textsc{Keribin, C.}, \textsc{Brault, V.}, \textsc{Celeux, G.} \&
  \textsc{Govaert, G.} (2015).
\newblock Estimation and selection for the latent block model on categorical
  data.
\newblock \textit{Statistics and Computing} \textbf{25}, 1201--1216.

\bibitem[{Korwar \& Hollander(1973)}]{kor-hol:con}
\textsc{Korwar, R.~M.} \& \textsc{Hollander, M.} (1973).
\newblock Contributions to the theory of {D}irichlet processes.
\newblock \textit{Annals of Probability} \textbf{1}, 705--711.

\bibitem[{Lange(1999)}]{Lange99}
\textsc{Lange, K.} (1999).
\newblock \textit{Numerical Analysis for Statisticians}.
\newblock New York: Springer-Verlag.

\bibitem[{Lau \& Green(2007)}]{lau-gre:bay}
\textsc{Lau, J.~W.} \& \textsc{Green, P.~J.} (2007).
\newblock Bayesian model-based clustering procedures.
\newblock \textit{Journal of Computational and Graphical Statistics}
  \textbf{16}, 526--558.

\bibitem[{Lee \& Robert(2016)}]{lee:robert:2016}
\textsc{Lee, J.} \& \textsc{Robert, C.} (2016).
\newblock Importance sampling schemes for evidence approximation in mixture
  models.
\newblock \textit{Bayesian Analysis} \textbf{11}, 573--597.

\bibitem[{Lee et~al.(2009)Lee, Marin, Mengersen \&
  Robert}]{lee:marin:mengersen:robert:2008}
\textsc{Lee, K.}, \textsc{Marin, J.-M.}, \textsc{Mengersen, K.} \&
  \textsc{Robert, C.} (2009).
\newblock {B}ayesian inference on mixtures of distributions.
\newblock In \textit{Perspectives in Mathematical Sciences I: Probability and
  Statistics}, N.~N. Sastry, M.~Delampady \& B.~Rajeev, eds. Singapore: World
  Scientific, pp. 165--202.

\bibitem[{Lee \& McLachlan(2013)}]{lee-mcl:emm}
\textsc{Lee, S.~X.} \& \textsc{McLachlan, G.~J.} (2013).
\newblock {EMMIX}uskew: An {R} package for fitting mixtures of multivariate
  skew t-distributions via the {EM} algorithm.
\newblock \textit{Journal of Statistical Software} \textbf{55}, 1--22.

\bibitem[{Liverani et~al.(2013)Liverani, Hastie, Papathomas \&
  Richardson}]{Mix:Liverani2013}
\textsc{Liverani, S.}, \textsc{Hastie, D.~I.}, \textsc{Papathomas, M.} \&
  \textsc{Richardson, S.} (2013).
\newblock {PReMiuM}: An {R} package for profile regression mixture models using
  {D}irichlet processes.
\newblock Preprint, arXiv:1303.2836.

\bibitem[{Malsiner-Walli et~al.(2016)Malsiner-Walli, Fr{\"u}hwirth-Schnatter \&
  Gr{\"u}n}]{MWFSG16}
\textsc{Malsiner-Walli, G.}, \textsc{Fr{\"u}hwirth-Schnatter, S.} \&
  \textsc{Gr{\"u}n, B.} (2016).
\newblock Model-based clustering based on sparse finite {G}aussian mixtures.
\newblock \textit{Statistics and Computing} \textbf{26}, 303--324.

\bibitem[{{Malsiner-Walli} et~al.(2017){Malsiner-Walli},
  Fr{\"u}hwirth-Schnatter \& Gr{\"u}n}]{mal-etal:ide}
\textsc{{Malsiner-Walli}, G.}, \textsc{Fr{\"u}hwirth-Schnatter, S.} \&
  \textsc{Gr{\"u}n, B.} (2017).
\newblock Identifying mixtures of mixtures using {B}ayesian estimation.
\newblock \textit{Journal of Computational and Graphical Statistics}
  \textbf{26}, 285--295.

\bibitem[{Marin et~al.(2005)Marin, Mengersen \&
  Robert}]{marin:mengersen:robert:2004}
\textsc{Marin, J.-M.}, \textsc{Mengersen, K.} \& \textsc{Robert, C.} (2005).
\newblock {B}ayesian modelling and inference on mixtures of distributions.
\newblock In \textit{Handbook of Statistics}, C.~R. Rao \& D.~Dey, eds.,
  vol.~25. New York: Springer-Verlag, pp. 459--507.

\bibitem[{McLachlan(1987)}]{MCLachlan87}
\textsc{McLachlan, G.} (1987).
\newblock On bootstrapping the likelihood ratio test statistic for the number
  of components in a normal mixture.
\newblock \textit{Applied Statistics} \textbf{36}, 318--324.

\bibitem[{McLachlan \& Peel(2000)}]{McLPeel}
\textsc{McLachlan, G.} \& \textsc{Peel, D.} (2000).
\newblock \textit{Finite Mixture Models}.
\newblock New York: Wiley.

\bibitem[{Meng \& Wong(1996)}]{meng:wong:1996}
\textsc{Meng, X.} \& \textsc{Wong, W.} (1996).
\newblock Simulating ratios of normalizing constants via a simple identity: A
  theoretical exploration.
\newblock \textit{{S}tatistica Sinica} \textbf{6}, 831--860.

\bibitem[{Miller \& Harrison(2018)}]{mil-har:mix}
\textsc{Miller, J.~W.} \& \textsc{Harrison, M.~T.} (2018).
\newblock Mixture models with a prior on the number of components.
\newblock \textit{Journal of the American Statistical Association}
  \textbf{113}, 340--356.

\bibitem[{Molitor et~al.(2010)Molitor, Papathomas, Jerrett \&
  Richardson}]{mol-etal:bay}
\textsc{Molitor, J.}, \textsc{Papathomas, M.}, \textsc{Jerrett, M.} \&
  \textsc{Richardson, S.} (2010).
\newblock Bayesian profile regression with an application to the {N}ational
  {S}urvey of {C}hildren's {H}ealth.
\newblock \textit{Biostatistics} \textbf{11}, 484--498.

\bibitem[{M{\"u}ller \& Mitra(2013)}]{mue-mit:bay}
\textsc{M{\"u}ller, P.} \& \textsc{Mitra, R.} (2013).
\newblock Bayesian nonparametric inference -- why and how.
\newblock \textit{Bayesian Analysis} \textbf{8}, 269--360.

\bibitem[{Nadif \& Govaert(1998)}]{Nadif}
\textsc{Nadif, M.} \& \textsc{Govaert, G.} (1998).
\newblock Clustering for binary data and mixture models -- choice of the model.
\newblock \textit{Applied Stochastic Models and Data Analysis} \textbf{13},
  269--278.

\bibitem[{Neal(1999)}]{neal:1999}
\textsc{Neal, R.} (1999).
\newblock Erroneous results in ``{M}arginal likelihood from the {G}ibbs
  output''.
\newblock Tech. rep., University of Toronto.

\bibitem[{Nobile(2004)}]{nob:pos}
\textsc{Nobile, A.} (2004).
\newblock On the posterior distribution of the number of components in a finite
  mixture.
\newblock \textit{Annals of Statistics} \textbf{32}, 2044--2073.

\bibitem[{Nobile \& Fearnside(2007)}]{nob-fea:bay}
\textsc{Nobile, A.} \& \textsc{Fearnside, A.} (2007).
\newblock Bayesian finite mixtures with an unknown number of components: The
  allocation sampler.
\newblock \textit{Statistics and Computing} \textbf{17}, 147--162.

\bibitem[{Pitman \& Yor(1997)}]{pit-yor:two}
\textsc{Pitman, J.} \& \textsc{Yor, M.} (1997).
\newblock The two-parameter {P}oisson-{D}irichlet distribution derived from a
  stable subordinator.
\newblock \textit{Annals of Probability} \textbf{25}, 855--900.

\bibitem[{Rau et~al.(2015)Rau, Maugis-Rabusseau, Martin-Magniette \&
  Celeux}]{RMMC15}
\textsc{Rau, A.}, \textsc{Maugis-Rabusseau, C.}, \textsc{Martin-Magniette,
  M.-L.} \& \textsc{Celeux, G.} (2015).
\newblock Co-expression analysis of high-throughput transcriptome sequencing
  data with {P}oisson mixture models.
\newblock \textit{Bioinformatics} \textbf{31}, 1420--1427.

\bibitem[{Richardson \& Green(1997)}]{richardson:green:1997}
\textsc{Richardson, S.} \& \textsc{Green, P.} (1997).
\newblock On {B}ayesian analysis of mixtures with an unknown number of
  components (with discussion).
\newblock \textit{Journal of the Royal Statistical Society, Series B}
  \textbf{59}, 731--792.

\bibitem[{Rissanen(2012)}]{riss}
\textsc{Rissanen, J.} (2012).
\newblock \textit{Optimal Estimation of Parameters}.
\newblock Cambrdige: Cambridge University Press.

\bibitem[{Robert(2007)}]{robert:2007}
\textsc{Robert, C.~P.} (2007).
\newblock \textit{The {B}ayesian Choice}.
\newblock New York: Springer.

\bibitem[{Robert \& Casella(2004)}]{robert:casella:2004}
\textsc{Robert, C.~P.} \& \textsc{Casella, G.} (2004).
\newblock \textit{{M}onte {C}arlo {S}tatistical {M}ethods}.
\newblock New York: Springer-Verlag, 2nd ed.

\bibitem[{Roeder(1990)}]{roe:den}
\textsc{Roeder, K.} (1990).
\newblock Density estimation with confidence sets exemplified by superclusters
  and voids in galaxies.
\newblock \textit{Journal of the American Statistical Association} \textbf{85},
  617--624.

\bibitem[{Roeder \& Wasserman(1997)}]{Roeder97}
\textsc{Roeder, K.} \& \textsc{Wasserman, L.} (1997).
\newblock Practical {B}ayesian density estimation using mixtures of normals.
\newblock \textit{Journal of the American Statistical Association} \textbf{92},
  894--902.

\bibitem[{Rousseau \& Mengersen(2011)}]{mengersen:rousseau:2011}
\textsc{Rousseau, J.} \& \textsc{Mengersen, K.} (2011).
\newblock Asymptotic behaviour of the posterior distribution in overfitted
  mixture models.
\newblock \textit{Journal of the Royal Statistical Society, Series B}
  \textbf{73}, 689--710.

\bibitem[{Schwartz(1965)}]{Schwartz}
\textsc{Schwartz, L.} (1965).
\newblock On {B}ayes procedures.
\newblock \textit{Zeitschrift f\"ur Wahrscheinlichkeitstheorie und verwandte
  Gebiete} \textbf{4}, 10--26.

\bibitem[{Schwarz(1978)}]{Schwarz}
\textsc{Schwarz, G.} (1978).
\newblock Estimating the dimension of a model.
\newblock \textit{Annals of Statistics} \textbf{6}, 461--464.

\bibitem[{Sethuraman(1994)}]{set:con}
\textsc{Sethuraman, J.} (1994).
\newblock A constructive definition of {D}irichlet priors.
\newblock \textit{Statistica Sinica} \textbf{4}, 639--650.

\bibitem[{Skilling(2007)}]{skilling:2007}
\textsc{Skilling, J.} (2007).
\newblock Nested sampling for {B}ayesian computations.
\newblock \textit{Bayesian Analysis} \textbf{1}, 833--859.

\bibitem[{Spiegelhalter et~al.(2002)Spiegelhalter, Best, Carlin \& {van der
  Linde}}]{spi-etal:baydic}
\textsc{Spiegelhalter, D.~J.}, \textsc{Best, N.~G.}, \textsc{Carlin, B.~P.} \&
  \textsc{{van der Linde}, A.} (2002).
\newblock Bayesian measures of model complexity and fit.
\newblock \textit{Journal of the Royal Statistical Society, Series B}
  \textbf{64}, 583--639.

\bibitem[{Tessier et~al.(2006)Tessier, Schoenauer, Biernacki, Celeux \&
  Govaert}]{TSBCG06}
\textsc{Tessier, D.}, \textsc{Schoenauer, M.}, \textsc{Biernacki, C.},
  \textsc{Celeux, G.} \& \textsc{Govaert, G.} (2006).
\newblock Evolutionary latent class clustering of qualitative data.
\newblock Tech. Rep. RR-6082, INRIA.

\bibitem[{{van H}avre et~al.(2015){van H}avre, White, Rousseau \&
  Mengersen}]{van-etal:ove}
\textsc{{van H}avre, Z.}, \textsc{White, N.}, \textsc{Rousseau, J.} \&
  \textsc{Mengersen, K.} (2015).
\newblock Overfitting {B}ayesian mixture models with an unknown number of
  components.
\newblock \textit{{PL}o{S ONE}} \textbf{10}, e0131739.

\bibitem[{Wade \& Gharhamani(2018)}]{wad-gha:bay}
\textsc{Wade, S.} \& \textsc{Gharhamani, Z.} (2018).
\newblock Bayesian cluster analysis: {P}oint estimation and credible balls
  (with discussion).
\newblock \textit{Bayesian Analysis} \textbf{13}, 559--626.

\bibitem[{Wallace \& Freeman(1987)}]{Wallace87}
\textsc{Wallace, C.} \& \textsc{Freeman, P.} (1987).
\newblock Estimation and inference via compact coding.
\newblock \textit{Journal of the Royal Statistical Society, Series B}
  \textbf{49}, 241--252.

\bibitem[{Watanabe(2009)}]{Watanabe2009}
\textsc{Watanabe, S.} (2009).
\newblock \textit{Algebraic Geometry and Statistical Learning Theory}.
\newblock Cambridge: Cambridge University Press.

\bibitem[{Wyse et~al.(2017)Wyse, Friel \& Latouche}]{WFL15}
\textsc{Wyse, J.}, \textsc{Friel, N.} \& \textsc{Latouche, P.} (2017).
\newblock Inferring structure in bipartite networks using the latent block
  model and exact {ICL}.
\newblock \textit{Network Science} \textbf{5}, 45--69.

\bibitem[{Yang(2005)}]{Yang05}
\textsc{Yang, Y.} (2005).
\newblock Can the strengths of {AIC} and {BIC} be shared? {A} conflict between
  model identification and regression estimation.
\newblock \textit{Biometrika} \textbf{92}, 937--950.

\bibitem[{Zeng \& Cheung(2014)}]{Zeng14}
\textsc{Zeng, H.} \& \textsc{Cheung, Y.-M.} (2014).
\newblock Learning a mixture model for clustering with the completed likelihood
  minimum message length criterion.
\newblock \textit{Pattern Recognition} \textbf{47}, 2011--2030.

\bibitem[{Zhang et~al.(2004)Zhang, Chan, Wu \& Chen}]{zha-etal:lea}
\textsc{Zhang, Z.}, \textsc{Chan, K.~L.}, \textsc{Wu, Y.} \& \textsc{Chen, C.}
  (2004).
\newblock Learning a multivariate {G}aussian mixture model with the reversible
  jump {MCMC} algorithm.
\newblock \textit{Statistics and Computing} \textbf{14}, 343--355.

\end{thebibliography}
\end{document}